\documentclass[a4paper,11pt]{article}
\pdfoutput=1 

\usepackage{jheppub} 

\usepackage[T1]{fontenc} 
\usepackage[usenames,dvipsnames]{xcolor}
\usepackage{feynmp-auto}

\newcommand{\half}{\frac{1}{2}}
\newcommand{\la}[1] {\left\langle #1 \right\rvert}
\newcommand{\ls}[1] {\left\lbrack #1 \bf \right\rvert}
\newcommand{\ra}[1] {\left\lvert #1 \right\rangle}
\newcommand{\rs}[1] {\left\lvert #1 \bf \right\rbrack}
\newcommand{\da}[1] {\left\langle #1 \right\rangle}
\newcommand{\ds}[1] {\left\lbrack #1 \bf \right\rbrack}
\newcommand{\dc}[1] {\left\{ #1 \bf \right\}}

\usepackage{verbatim}

\usepackage{amstext} 
\usepackage{array}   
\newcolumntype{L}{>{$}l<{$}} 

\title{Constructing \texorpdfstring{$\mathcal{N}=4$}{N=4} Coulomb Branch Superamplitudes}

\author[a,b]{Aidan Herderschee,}
\author[a]{Seth Koren,}
\author[a]{and Timothy Trott}
\affiliation[a]{Department of Physics, University of California, \\
Santa Barbara, CA 93106, U.S.A.}

\affiliation[b]{Leinweber Center for Theoretical Physics, \\
Randall Laboratory of Physics, Department of Physics, \\
University of Michigan, Ann Arbor, MI 48109, USA}

\emailAdd{aidanh@umich.edu}
\emailAdd{koren@physics.ucsb.edu}
\emailAdd{ttrott@physics.ucsb.edu}

\abstract{We study scattering amplitudes of massive BPS states on the Coulomb branch of $4d$ $\mathcal{N}=4$ super-Yang-Mills, utilising a little group covariant on-shell superspace for massive particles. Super-BCFW recursion for massive amplitudes is constructed and its validity is proven for all Coulomb branch superamplitudes. We then determine the exact three-particle superamplitudes for massive states. These ingredients allow us to explicitly compute the four- and five-particle superamplitudes, which is the first non-trivial usage of BCFW recursion for amplitudes with entirely massive external states. The manifest little group covariance helps clarify both the role of special kinematic properties of BPS states and the organizational structures of the superamplitudes.
}

\begin{document}

\maketitle

\section{Introduction}

The most powerful on-shell properties are to be found with maximal $\mathcal{N}=4$ supersymmetry (at least for non-gravitational theories). Although a highly idealised model of QCD, numerous hidden structures beyond the maximal, rigid supersymmetry have been uncovered and their role in nature remains to be ascertained. Some particular highlights include the computation of tree amplitudes at strong coupling by holography \cite{Alday:2007hr}, the duality of planar (large number of colours) amplitudes with Wilson loops \cite{Drummond:2007aua,Henn:2009bd,CaronHuot:2010ek,Mason:2010yk,Adamo:2011dq}, the discovery of dual (super)conformal symmetry (in addition to regular spacetime superconformal symmetry) \cite{Drummond:2008vq,Berkovits:2008ic}, Yangian symmetry and integrable structure \cite{Drummond:2009fd}, constructibility of tree \cite{Drummond:2008cr} amplitudes by BCFW recursion \cite{Britto:2004ap,Britto:2005fq}, loop integrands by on-shell diagrams and full constructibility from leading singularities \cite{ArkaniHamed:2010gh,ArkaniHamed:2010kv} and the interpretation of amplitudes as volumes of polytopes \cite{Hodges:2009hk,Arkani-Hamed:2013jha}. Most of this work has focused on the origin of the moduli space, where the states are all massless and the theory is conformal. 

The structure of amplitudes of massive particles with $\mathcal{N}=4$ supersymmetry has received comparatively little attention. These nevertheless provide a further testing ground of the special symmetries and properties listed above and the extent to which they are deformed but not destroyed by Higgsing. Previous studies of massive amplitudes on the Coulomb branch have been made in \cite{Boels:2010mj,Craig:2011ws,Kiermaier:2011cr,Elvang:2011ub}, where a gamut of methods including soft limits, supersymmetric on-shell recursion and solutions to the supersymmetric Ward identities (SWIs) were proposed and used to compute some simple examples. Subsequently, some $4d$ tree-level amplitudes and loop integrands have been obtained by dimensional reduction from superamplitudes of the $6d$ $\mathcal{N}=(1,1)$ SYM theory, for which dual conformal invariance has been established, despite the absence of conformality \cite{Dennen:2009vk,Dennen:2010dh,Bern:2010qa,Huang:2011um,Plefka:2014fta}. However, a general procedure for explicitly constructing amplitudes beyond the fewest leg examples was not developed. More recently, a CHY \cite{Cachazo:2013gna} formula for all $6d$ $\mathcal{N}=(1,1)$ massless amplitudes was found and reduced to give a general formula for all $4d$ massive $\mathcal{N}=4$ tree amplitudes \cite{Cachazo:2018}, from which a few examples were extracted (a new proposal was recently made in \cite{Geyer:2018xgb}). Partial use of the massive spinor helicity formalism discussed here was made to extract some simple examples of amplitudes contained within the general formula. Nevertheless, much of the structure of these amplitudes thus far remains unexplored. We will review this subject more thoroughly in Section \ref{MassiveAmplitudes}.

To proceed onto the Coulomb branch, we first discuss an on-shell superspace for massive BPS vector multiplets. Purely through the use of on-shell properties and maximal rigid supersymmetry, we construct the unique elementary three particle superamplitudes of the theory. These superamplitudes of massive legs have `nonlocal' kinematic denominators analogous to that present in massless (S)YM, despite this feature not being present in any of the component amplitudes. This arises as a result of the special complex kinematics of the BPS states and suggests that the massive amplitudes share in the special constructibility properties of massless gauge theory. We confirm this by formulating a massive super-BCFW shift and proving the constructibility of all Coulomb branch tree amplitudes under it. Using this to fuse the four particle superamplitude from a single factorization channel between on-shell three-leg superamplitudes, we are able to explicitly locate the second pole of the four-point superamplitude as coming from the singular overlap of the two special kinematic configurations on either side of the factorization channel. 

The establishment of super-BCFW for massive legs allows for the systematic computation of relatively compact expressions for massive superamplitudes. To illustrate this, we explicitly write down the five particle superamplitude for all-massive legs, which is the first non-trivial usage of on-shell recursion to construct an amplitude of fully massive external states. The way in which the massless sectors of helicity violation combine together when the states are massive is also shown.

This work is partnered with a companion paper \cite{HKT:2018a} that discusses the on-shell properties of supersymmetric theories with massive particles (mostly with $\mathcal{N}=1$ supersymmetry). This makes use of the adaptation of helicity spinors to describe the kinematics of massive particles made in \cite{Arkani-Hamed:2017jhn} with manifest little group covariance. 

This paper takes the following steps toward elucidating the structure of massive amplitudes in $\mathcal{N}=4$ SYM. We firstly review, in Section \ref{sec:Review}, the representation theory of massive particles pertinent to the Coulomb branch of $\mathcal{N}=4$. In Section \ref{N=4superspace}, we introduce the `non-chiral' superspace in which the superamplitudes are naturally formulated and explain the representation of BPS states (here massive elementary vector multiplets) in on-shell superspace. In order to construct higher-leg amplitudes, we implement BCFW recursion for massive superamplitudes in Section \ref{sec:massiveBCFW} and establish that all Coulomb branch amplitudes are constructible in this manner. In Section \ref{MassiveAmplitudes}, we commence the calculation of massive scattering amplitudes. We find the three-particle superamplitudes in subsection \ref{Sec:3leg}, which features a `special kinematics' of BPS states resembling that of massless particles with complex momenta, as well as a surprising `nonlocality' in their superamplitudes. This enables us to recursively construct the four-leg superamplitude in subsection \ref{sec:4leg} (with some computational details shunted to Appendix \ref{Appendix4leg}). In subsection \ref{sec:5ptCoulomb}, after a discussion of the supersymmetric `band structure', we are able to use the same technique to find the five particle superamplitude for all-massive states. We then conclude. In Appendix \ref{sec:projecting} we make some comments about projecting Coulomb branch superamplitudes down to Yang-Mills theories with massive particles with fewer supersymmetries.

\section{On-shell superfields for massive particles}\label{sec:Review}

In \cite{HKT:2018a} we construct on-shell superspaces for massive supermultiplets that are covariant in the $SU(2)$ little group, recently introduced into helicity spinors in \cite{Arkani-Hamed:2017jhn}. We here briefly summarize the important results and refer the reader to \cite{HKT:2018a} for further details, especially its appendix of conventions and identities.

For $\mathcal{N}$-extended SUSY, the supercharges carried by leg $i$, $Q_{i,\alpha A}$ and $ Q^{\dagger A}_{i,\dot{\alpha}}$, satisfy the commutation relations
\begin{align}
\{ Q_{i,\alpha A},Q_{i,\beta B}\}&=Z_{i,AB} \epsilon_{\alpha\beta}\qquad
\{ Q^{\dagger A}_{i,\dot{\alpha}},Q^{\dagger B}_{i,\dot{\beta}}\}=-Z^{AB}_i\epsilon_{\dot{\alpha}\dot{\beta}}\nonumber\\
&\{ Q_{i,\alpha  A},Q^{\dagger B}_{i,\dot{\beta}}\}=-2 \delta_{A}^{B}(\sigma^{\mu}_{\alpha\dot{\beta}})P_{i,\mu}\label{FullSUSYalgebra}
\end{align}
where $P_{i,\mu}$ is the momentum and $Z_{i,AB}$ is the central charge, satisfying $Z_{i,AB}=-Z_{i,BA}=-(Z^{AB}_i)^*$. The labels $A$ and $B$ are $R$-indices. On-shell, little group covariant supersymmetry generators are defined for each leg by projecting the supercharges onto the spinors of a given particle
\begin{equation}
q^I_{i,A} = \frac{-1}{\sqrt{2}m_i} \ds{i^I Q_{i,A}}, \qquad q^{\dagger A}_{i,I} = \frac{1}{\sqrt{2}m_i} \da{i_I Q^{\dagger A}_{i}},\label{supercharges}
\end{equation}
\noindent which satisfy the anticommutation relations
\begin{equation}
\dc{q_{i,A}^{I},q^{\dagger J,B}_{i}}=-\epsilon^{IJ}\delta_{A}^{B}, \qquad \dc{q_{i,A}^{I}, q_{i,B}^{J}}=-\epsilon^{IJ}\frac{Z_{i,AB}}{2m_i}, \qquad \dc{q_{i}^{\dagger I,A}, q_{i}^{\dagger J,B} }=\epsilon^{IJ}\frac{Z_i^{AB}}{2m_i}.\label{SUSYalgebra}
\end{equation}
The index $I$ denotes massive $SU(2)$ little group component while $m_i$ is the mass of the leg. For the simplest case, which will be considered here, $Z_{i,AB}=Z_i\,\Omega_{AB}$ for all $i$, where $Z_i\in\mathbb{R}$ while $\Omega_{AB}=-\Omega_{BA}$ is a symplectic $2$-form
\begin{equation}\label{eqn:centralcharge}
\Omega_{AB}=\begin{bmatrix}
 0 & -I \\ 
 I & 0
\end{bmatrix}.
\end{equation}
The case $|Z_i|=2m_i$ is the special BPS limit and will be relevant for states on the Coulomb branch. For these representations, half of the supercharges are eliminated through the reality constraint
\begin{equation}
q_{i,IA}=\frac{-1}{2m_i}Z_{i,AB}q^{\dagger B}_{i,I}.
\label{bpslimit}
\end{equation}

\noindent The phase of $Z$ may be absorbed into a redefinition of the supercharges $q_i$ and $q_i^\dagger$. This condition again preserves the supersymmetry algebra. BPS states are annihilated by the combination $q^{IA}_{i}\pm q^{\dagger IA}_{i}$ (the sign is determined by the sign of $Z_i$). For non-BPS representations with a central charge, linear combinations of supercharges may be found that will satisfy the algebra (\ref{SUSYalgebra}) with $Z_{i,AB}=0$. The representation theory of these states is therefore unaffected by the existence of a central charge.

The explicit $SU(\mathcal{N})$ automorphism symmetry of the SUSY algebra is broken to $USp(\mathcal{N})$ by the central charge of these massive single particle states, which is exactly the massive $R$-symmetry group expected for a theory with half of the number of supersymmetries. A BPS state in $\mathcal{N}$-SUSY may be represented as a massive non-BPS state of $\mathcal{N}/2$-SUSY. For the simplest symmetry breaking pattern of the $\mathcal{N}=4$ SYM Coloumb branch, the massless $SU(4)$ $R$-symmetry is broken to $USp(4)$ when the central charge is generated.  

From (\ref{SUSYalgebra}), the massive supersymmetry algebra is that of $N$ fermionic oscillators, where $N=2\mathcal{N}$ if the representation is not BPS, but can be reduced by up to a factor of $1/2$ if shortened. Supermultiplets may be represented as coherent states which are eigenstates of $N$ `lowering operators'. To build these states we introduce Grassmann variables which transform as fundamental spinors of the little group of each particle $\eta^A_{i,I}$, as well as their conjugates $\eta^{\dagger I}_{i,A}$. The $R$-index on the Grassmann variables is truncated for $1/2$-BPS states to denote some subset of the $\mathcal{N}/2$ supersymmetries that do not leave the state invariant. We will use the fact that BPS states of $\mathcal{N}=4$ obey the same algebra as the non-BPS state of $\mathcal{N}=2$, which simplifies its construction.

To ensure little group covariance, we choose all of the $q^{\dagger A}_{i,I}$ as the lowering operators. An entire supermultiplet may be encoded as a coherent state
\begin{equation}
\la{\eta_i} = \la{\Omega} e^{q^{ I}_{i,A}\eta^A_{i,I} } 
\end{equation} 
where $\eta^A_{i,I}$ are anticommuting Grassmann algebra generators and $\la{\Omega}$ is the Clifford vaccum annihilated by $q^{\dagger A}_{i,I}$. These are eigenstates of the annihilation operators, satisfying $\la{\eta_i}q^{\dagger}_{i,I}= \la{\eta_i}(-\eta_{i,I})$. The action of the supercharges on the coherent states may be represented as 
\begin{equation} \label{eqn:ampspinorsupercharges}
\begin{split}
q^{\dagger A}_{i, I}=-\eta_{i, I}^A&\qquad q^{ I}_{i, A}=-\frac{\partial}{\partial\eta_{i,I}^A}.
\end{split}
\end{equation}

Supersymmetry transformations generated by $q$ and $q^\dagger$ act simply on these coherent states:
\begin{equation} \label{eqn:susyxforms}
\la{\eta} e^{i\xi^{\dagger I}_{A} q_{i,I}^{\dagger A}} =  e^{-i\xi^{\dagger I}_{A} \eta_I^{A}}\la{\eta}, \qquad \la{\eta} e^{-i\xi^A_{I} q_{i,A}^{I}} = \la{\eta + i\xi}.
\end{equation}
Here, $\xi_I^{A}=\ds{\theta^A i_I}$ and $\xi^{\dagger I}_{A}=\da{\theta_A i^I}$ parameterise the supersymmetry transformation projected onto the spinors of leg $i$ of the appropriate chirality, for some Grassmann spinors $\ls{\theta^A}$ and $\la{\theta^A}$. The action of the supercharges encoded in (\ref{eqn:susyxforms}) give the supersymmetric Ward identities (SWIs) relating the components.

Only elementary massive vector multiplets will be of interest to us in our investigation of scattering amplitudes on the Coulomb branch of $\mathcal{N}=4$. These are half-BPS, which are equivalent to long $\mathcal{N}=2$ vector multiplets. Expanding the $\mathcal{N}=2$ coherent state gives the superfield 
\begin{equation}
\mathcal{W} =\phi+\eta^a_I\psi^I_a-\half \eta^a_I \eta^b_J (\epsilon^{IJ} \phi_{(ab)} + \epsilon_{ab} W^{(IJ)})+\frac{1}{3}\epsilon_{bc}\eta_I^b\eta_{J}^c\eta^{Ja}\tilde{\psi}_a^I +\eta_1^1\eta_1^2\eta_2^1\eta_2^2\tilde{\phi}
\label{LongMultNtwo},
\end{equation}
See \cite{HKT:2018a} for details. The $R$-indices $a,b,c$ are those of the $SU(2)_R$ of the $\mathcal{N}=2$ SUSY algebra. The states $\phi$, $\tilde{\phi}$ and $\phi_{(ab)}$ represent $5$ scalar quanta, $\psi_a^I$ and $\tilde{\psi}_a^I$ represent the degrees of freedom of two Dirac fermions, while $W^{(IJ)}$ represents the spin triplet of massive vector states. This superfield and its massless limit will be discussed further in Section \ref{BPS}.

\section{On-Shell Superspace for the \texorpdfstring{$\mathcal{N}=4$}{N=4} Coulomb branch}\label{N=4superspace}

\subsection{Non-Chiral Superspace}

The massless supermultiplet of $\mathcal{N}=4$ at the origin of moduli space is commonly constructed in the `chiral superspace' in which it is represented as a coherent state of $\eta^A$ for $SU(4)$ index $A$ (e.g. see for review \cite{elvang2015scattering}). These carry massless $U(1)$ helicity weights. This leads to a superfield\footnote{We express this in the form of \cite{elvang2015scattering}, defining the phases of the states to be those necessary to produce this from the action of $q_{i,A}$ on the Clifford vaccuum.}
\begin{equation}
G^+=g^{+}+\eta^{A}\lambda^+_{A}-\frac{1}{2}\eta^{A}\eta^{B}S_{AB}-\frac{1}{6}\eta^{A}\eta^{B}\eta^{C}\lambda^-_{ABC}+\eta^{1}\eta^{2}\eta^{3}\eta^{4}g^{-},
\label{masslessN4}
\end{equation}
where the superscript on the superfield labels the helicity of the supermultiplet. This contains the gluon $g^\pm$, four chiral gauginos with positive and negative helicities $\lambda^+_A$ and $\lambda^-_{ABC}$ respectively (the latter is totally antisymmetric in its $R$-indices and has only four independent components) and three complex scalars $S_{AB}$ satisfying self-duality $S_{AB}=\frac{1}{2}\epsilon_{ABCD}S^{*CD}$. 

However, we will find in what follows that for the supercharges to be represented as homogeneously multiplicative or derivative on the superfields in the presence of massive BPS states, we are led to construct the massless multiplets in the `non-chiral superspace', introduced in \cite{Huang:2011um}. To find the non-chiral superspace representation of the massless multiplet, we may perform a `half-Fourier transform' from $\eta^3,\eta^4$ to $\eta^\dagger_3, \eta^\dagger_4$. This construction is natural from the perspective of the dimensional reduction of $6$d $\mathcal{N}=(1,1)$ SYM to $4$d $\mathcal{N}=4$ SYM, as used in \cite{Huang:2011um,Plefka:2014fta,Cachazo:2018} (also see \cite{Heydeman:2017yww} for developments of on-shell superspaces for similar $4d$ and $6d$ theories on brane world-volumes). The massless superfield will now be a coherent state expanded in $\eta^a$, for $a = 1,2$, and $\eta^\dagger_m$, for $m=3,4$. The manifest massless $R$-symmetry is thus reduced to $SU(2)\times SU(2)$, although the multiplet remains $SU(2,2)$ invariant. This form was used for the non-chiral superspace of \cite{Plefka:2014fta}.

However, this $SU(2)\times SU(2)$ is not a subgroup of the unbroken $R$-symmetry group $USp(4)$ (or $USp(2,2)$ after the half-Fourier transform), so will be broken in the superamplitudes on the Coulomb branch. Instead, as will become clearer in our discussion of BPS multiplets below, we will find it more useful to manifest a representation of a $U(2)\leq USp(4)$, under which the fundamental $USp(4)$ vector decomposes as $\mathbf{4}\rightarrow \mathbf{2} \oplus \bar{\mathbf{2}}$. Then $\eta^a$ and $\tilde{\eta}^{\dagger a}=\eta^\dagger_{a+2}$ both transform in the $\mathbf{2}$ representation of this $U(2)$ subgroup. In this notation, heights of the $R$-indices on the states in (\ref{masslessN4}) are reversed for $A=3,4$ to show explicit $U(2)$ invariance of the superfield. The supermultiplet in the non-chiral superspace, first in the form of \cite{Plefka:2014fta} with the manifest broken $SU(2)\times SU(2)$ and second in the form with the (partially) manifest $U(2)$, is 
\begin{equation}
\begin{split}
G &= - \frac{1}{2}S^m_{\ m} + \eta^{\dagger}_{m}\lambda^{+m} + \frac{1}{2}\eta^{a}\lambda^{-\ m}_{am}  - \frac{1}{2} \eta_{a}\eta^{a} g^{-} + \half \eta^{\dagger}_{m}\eta^{\dagger m}g^{+} \\
&\quad\quad+\eta^{a}\eta^\dagger_{m}S^{m}_{\ a}+\half \eta^{\dagger}_{m} \eta^{\dagger m}\eta^{a}\lambda^+_{a}+\frac{1}{4} \eta_{a}\eta^{a} \eta^{\dagger}_{m}\lambda^{-m b}_{\ \ \ \ \ b} -\frac{1}{4}\eta^{\dagger}_{m}\eta^{\dagger m}\eta_{a}\eta^{a}S_b^{\ b}\\
&= S_{34} + (\tilde{\eta}^{\dagger 1}\lambda^+_{4}-\tilde{\eta}^{\dagger 2}\lambda^+_{3}) + \eta^{a}\lambda^-_{a34} - \eta^{1}\eta^{2} g^{-} + \tilde{\eta}^{\dagger 1}\tilde{\eta}^{\dagger 2} g^{+} \\
&\quad\quad+\eta^{a}(\tilde{\eta}^{\dagger 1}S_{4\,a}-\tilde{\eta}^{\dagger 2}S_{3\,a})+\tilde{\eta}^{\dagger 1}\tilde{\eta}^{\dagger 2}\eta^{a}\lambda^+_{a} + \frac{1}{2} \eta^{a}\eta^{b}(\tilde{\eta}^{\dagger 2}\lambda^-_{ab3}-\tilde{\eta}^{\dagger 1}\lambda^-_{ab4}) + \tilde{\eta}^{\dagger 2}\tilde{\eta}^{\dagger 1}\eta^{1}\eta^{2}S_{12}.
\label{masslessmixed}
\end{split}
\end{equation}
The latter form will be henceforth assumed, although this will not actually be very important in what follows. In the former expression, index heights in each $SU(2)$ sector may be raised and lowered with the Levi-Civita symbol as usual. However, in the latter form, $G$ is charged under a $U(1)\leq U(2)$ subgroup. Each Grassmann variable carries a unit charge under a $U(1)$ subgroup, while the states are also charged such that each term above has an overall charge of $+2$ units. While possible to adjust the notation to make the $SU(2)\leq U(2)$ invariance manifest, we find that, in practice, the above form is clearest (these expressions are mostly useful for identifying extraction functions to find component amplitudes). 

The superfield (\ref{masslessmixed}) is the massless counterpart to the massive superfield in (\ref{LongMultNtwo}). The correspondance between the massless and massive on-shell superspace variables will be elaborated upon below.

The (complexified) $R$-symmetry generators for the $USp(2,2)$ on the non-chiral superspace are 
\begin{align}\label{MasslessR}
\mathbf{m}_b^a=\sum_i\Big(\tilde{\eta}^{\dagger a}_{i}\frac{\partial}{\partial\tilde{\eta}^{\dagger b}_{i}}+\eta^{a}_{i}\frac{\partial}{\partial\eta^{b}_{i}}-2\delta^a_b\Big)\qquad \mathbf{k}_{ab}=\sum_i\frac{\partial}{\partial\tilde{\eta}_{i}^{\dagger(a}}\frac{\partial}{\partial\eta^{b)}_{i}}\qquad \mathbf{p}^{ab}=\sum_i\tilde{\eta}^{\dagger (a}_{i}\eta^{b)}_{i}.
\end{align}
The symbols have been chosen to reflect the resemblance to the conformal group. All massless legs $i$ are summed over. The reader is referred to \cite{Plefka:2014fta} for a larger catalogue of symmetry generator representations for the massless superfields in the non-chiral superspace. 

Pure $\mathcal{N}=4$ super-Yang-Mills theory has a supersymmetry-preserving moduli space of vacua upon which the scalar components of the vector supermultiplets acquire a vev and spontaneously break the gauge theory to some smaller rank unbroken subgroup. We will generally consider the possibility of multiple breakings of the gauge group to factors of $\prod_k U(N_k)$. For simplicity, we will assume that the scalars' vevs are of the form $\langle S_{AB}\rangle=\oplus_kv_k\delta_{i_k}^{j_k}\Omega_{AB}$ for some $v_k\in\mathbb{R}$. Here, $i_k$ and $j_k$ are gauge indices of an unbroken $U(N_k)$ subgroup. This breaking pattern induces a central charge $Z_{AB}\propto\Omega_{AB}$ and modifies the SUSY algebra to the form discussed above. The $R$-symmetry in this case is broken to $USp(4)$, which corresponds to the simplest case in which there is only a single central charge. The vector superfields that become massive through this Higgsing are BPS states and are bifundamentals of two of the unbroken gauge group factors. Calling these $U(N_{k_a})\times U(N_{k_b})$, then their masses are $g|v_{k_a}-v_{k_b}|$, where $v_{k_a}$ and $v_{k_b}$ are the vevs that break the generators corresponding the the vector superfields. Conservation of the central charge then implies that, in any scattering process, the sum of the masses of the particles (states of positive central charge) must be equal to the sum of the masses of the antiparticles (states of negative central charge). This selection rule places an extra kinematic constraint upon the amplitudes.

\subsection{BPS States}\label{BPS}

In the BPS case, the supersymmetry generators satisfy the reality condition

\begin{equation}
P^{\dot{\alpha}\alpha}_iQ_{i,\alpha,A}=\frac{1}{2}Z_{i,AB}Q^{\dagger,B\dot{\alpha}}_i,\label{BPS0}
\end{equation}

\noindent which implies (\ref{bpslimit}) when the little group symmetry is made manifest. This reduces the effective number of left-handed fermionic generators from $\mathcal{N}$ to $\mathcal{N}/2$. We use these remaining $\mathcal{N}/2$ generators to construct `short' BPS supermultiplets that are equivalent to the `long' massive supermultiplets of unextended $\mathcal{N}/2$ supersymmetry. For the Coulomb branch of $\mathcal{N}=4$ SYM, the massive multiplets will all be short multiplets, which are equivalent to the $\mathcal{N}=2$ multiplet given in (\ref{LongMultNtwo}). 

There is a choice in how to represent the BPS SUSY algebra, which corresponds to a choice of raising and lowering operators for our supermultiplets. This affects the organization both of states and of superamplitudes in theories with BPS multiplets, such as $\mathcal{N}=4$ on the Coulomb branch. Given our intent, it seems natural that we should make the choice which preserves manifest little group covariance of our BPS states. The formulation of an $\mathcal{N} = 2$ theory with BPS multiplets may be understood analogously, so we focus predominantly here on what happens for $\mathcal{N}=4$. A similar on-shell superspace for $\mathcal{N}=8$ supergravity incorporating half-BPS black holes was recently constructed in \cite{Caron-Huot:2018ape}.

Firstly, on the BPS states, the supercharges satisfy $q_i^{\dagger I,A}=-q_i^{I,A}$. It is at this point that the breaking of the $R$-structure of the supercharges into the non-chiral form discussed above for massless representations becomes natural for describing the BPS states. For $\mathcal{N}=4$, after decomposing the supercharges into two separate pairs independently transforming under the $U(2)$ $R$-subgroup, the BPS condition equates supercharges of one doublet with the conjugates of the other, which are in the same $U(2)$ representation. The massive BPS on-shell superfield may then be expanded in two little group pairs of Grassmann variables $\eta^{a}_{I}$ (for $a\in \{ 1,2\}$), just as for the massive $\mathcal{N}=2$ superfield derived above. The supercharges are then represented on these as $q_{i, I}^{\dagger a}=-q_{i, I, a+2}=-\eta_{i, I}^a$ and $q_{i, a}^{I}=q_{i}^{\dagger I, a+2}=-\frac{\partial}{\partial\eta_{i, I}^a}$, for $a\in\{1,2\}$. 

The anti-BPS superfields consist of the $CP$ conjugate states of the BPS superfields. For the anti-BPS states, the same coherent state basis may be selected, although, as the central charge has the opposite sign, the anti-BPS condition involves a relative negative sign $q_i^{\dagger I a}=q_i^{I a}$. This leaves a relative negative sign in the representations of the supercharges on the superspace compared to the BPS states. States at level $n$ in the BPS superfield are conjugate to states at level $\mathcal{N}-n$ in the anti-BPS superfield.

While only a $U(2)$ subgroup of the $R$-symmetry is manifest on the BPS multiplets, the full $USp(4)$ is still respected by the superamplitudes. The (complexified) $USp(4)$ $R$-symmetry generators (or, more precisely, $USp(2,2)$) represented on massive superfields are 
\begin{align}\label{MassiveR}
\mathbf{m}_b^a=\sum_i\Big(\eta^a_{i,I}\frac{\partial}{\partial\eta^b_{i,I}}-2\delta^a_b\Big)\qquad \mathbf{k}_{ab}=\frac{1}{2}\sum_i\pm\frac{\partial}{\partial\eta_{i,I}^{(a}}\frac{\partial}{\partial\eta^{b),I}_{i}}\qquad \mathbf{p}^{ab}=\frac{1}{2}\sum_i\pm\eta^{(a}_{i,I}\eta^{b),I}_{i}.
\end{align}
See discussion of the representation theory of the symplectic groups in \cite{Gunaydin:1990ag}. The $(+)$ in the $\mathbf{k}$ and $\mathbf{p}$ generators is for BPS legs and the $(-)$ is for anti-BPS. Note that the little group index on the Grassmann derivative is raised and lowered by $-\epsilon$ rather than $\epsilon$, so e.g. $\rs{i^I}\frac{\partial}{\partial\eta_{i, I}^a}=\rs{i_I}\frac{\partial}{\partial\eta_{i}^{I,a}}$. The expressions for the generators on the massless legs in (\ref{MassiveR}) should be combined with those stated above to obtain the representation of the full superamplitude.

The massless limit of the BPS superfield in the form (\ref{LongMultNtwo}) produces the non-chiral representation of the massless superfield (\ref{masslessmixed}), which makes clearer why this representation is natural when formulating Coulomb branch superamplitudes. In this limit, the two supermultiplets are related as
\begin{center}
\setlength\extrarowheight{4pt}
\begin{tabular}{ |L||L|L|L|L|L|L| } 
 \hline
 \text{Massive} & \phi & \psi^{I}_{a} & W^{IJ} & \phi_{ab} & \tilde{\psi}^{I}_{a} & \tilde{\phi}\\ \hline 
 \text{Massless} & S_{34} & \lambda_{a}^-, \lambda_{a+2}^+ & g^{\pm}, S_{13}+S_{24} & S_{14}, S_{13}-S_{24}, S_{23} & \lambda_{a}^+, \lambda_{a+2}^- & S_{12} \\ 
 \hline
\end{tabular}
\end{center}
\noindent The massless limit of the $\mathcal{N}=4$ BPS superfield therefore amounts to breaking up the little group indices, as we are familiar with in the non-supersymmetric case. We here send $\eta^{A}_{-}\rightarrow \eta^{a}$,  $\eta^{A}_{+}\rightarrow \tilde{\eta}^{\dagger a}$ (where $A$ here is the $\mathcal{N}=2$ $R$-index used in (\ref{LongMultNtwo})). For the anti-BPS states, as a consequence of our definition of the massive superspace variables given above, the massless limit is modified to $\eta^{A}_{+}\rightarrow -\tilde{\eta}^{\dagger a}$, as is required from the inverse relations implied by (\ref{supercharges}). The $R$-symmetry generators (\ref{MassiveR}) clearly match onto (\ref{MasslessR}). The fact that our covariant representation of the BPS state reduces to a mixed representation of the massless coherent state with a scalar Clifford vacuum suggests that this mixed (or non-chiral) representation may be useful for representing amplitudes on the Coulomb branch of $\mathcal{N}=4$. Previous works have instead \cite{Craig:2011ws, Kiermaier:2011cr} implicitly worked with a massive representation that manifested an $SU(2)\times SU(2)$ subgroup\footnote{This is a distinct subgroup from the broken $SU(2)\times SU(2)$ mentioned in the discussion preceding (\ref{masslessmixed}).} of the $USp(4)$ $R$-symmetry in which the massive little group was obscured. This representation led to massive coherent states that appear similar to the first expression in (\ref{masslessmixed}), but 
with $R$-indices broken into pairs $(\eta^1,\eta^3)$ and $(\eta^2,\eta^4)$
and the vector's longitudinal mode replacing a single scalar. Similar tension in manifesting $R$-symmetries and little group symmetries in on-shell superspaces arises in $6d$ \cite{Dennen:2010dh}. Here we note that the BPS states $\mathcal{W}$ on the Coulomb branch are not self-conjugate (being eigenstates of the central charge) and so their massless limits are likewise complex. 

The choice of non-chiral coherent state for the massless fields combines with the coherent state bases for the BPS states to ensure that the total supercharges $Q^{\dagger a}$ and $Q_{a+2}$ act multiplicatively on the superamplitudes (while their conjugates act on each leg homogeneously as derivatives). The full supercharges are therefore represented as 
\begin{align}\label{eqn:coulombscharges}
\frac{1}{\sqrt{2}}Q_a=\rs{i_{I}}\frac{\partial}{\partial \eta_{i,I}^a}+\rs{j_{I}}\frac{\partial}{\partial \eta_{j,I}^a}+\rs{k}\frac{\partial}{\partial\eta_{k}^a},\qquad
&\frac{1}{\sqrt{2}}Q^{\dagger a}=-\ra{i^{I}}\eta_{i,I}^{a} -\ra{j^{I}}\eta_{j,I}^{a}+\ra{k}\eta_{k}^a,\nonumber\\
\frac{1}{\sqrt{2}}Q_{a+2}=\rs{i^{I}}\eta_{i,I}^{a} -\rs{j^{I}}\eta_{j,I}^{a}+\rs{k}\tilde{\eta}^{\dagger a}_{k},\qquad
&\frac{1}{\sqrt{2}}Q^{\dagger a+2}=\ra{i_{I}}\frac{\partial}{\partial \eta_{i,I}^{a}}-\ra{j_{I}}\frac{\partial}{\partial \eta_{j,I}^{a}}+\ra{k}\frac{\partial}{\partial\tilde{\eta}^{\dagger a}_{k}},
\end{align}
where legs labeled $i$, $j$ and $k$ respectively enumerate $\mathcal{W}$, $\overline{\mathcal{W}}$ and $G$ legs and are implicitly summed over here. With the supercharges in this homogeneous form, the SWIs should be simplified.

While not considered here for simplicity, it is also possible to consider further breakings of the $R$-symmetry on the $\mathcal{N}=4$ Coulomb branch, by moving the vevs of the other scalar components away from the origin of the moduli space. As operators acting on external legs of elementary vector supermultiplets, the central charge eigenvalues $Z_{i,AB}$ may always be $SU(4)$ $R$-rotated into a form $Z_{i,AB}=z_i\Omega_{AB}$ and the BPS condition is unchanged. See \cite{Fayet:1978ig,Osborn:1979tq,Fraser:1997nd} for discussion. However, if the $R$-symmetry is broken beyond $USp(4)$, this rotation is leg-dependent and the form of the supercharges represented on the full superamplitude (and hence the SWIs) will be more complicated. On-shell representations of BPS states with more complicated configurations of central charges were recently discussed in \cite{Caron-Huot:2018ape} in the context of $\mathcal{N}=8$ SUGRA for BPS black holes.

\subsection{Superamplitude preliminaries}

Our ultimate ambition is to construct an arbitrary $n$-point amplitude with both massless and massive external states, $\mathcal{A}_{n}(\mathcal{W}_{1},\mathcal{W}_{2}\ldots,\overline{\mathcal{W}}_{j},\overline{\mathcal{W}}_{j+1}\ldots, G_{k},G_{k+1},\ldots)$. As is conventional in discussions of scattering amplitudes in gauge theories, we will be henceforth implicitly describing colour-stripped partial amplitudes $\mathcal{A}_n[\mathcal{W}_{1},G_2,G_3\ldots\overline{\mathcal{W}}_{j},G_{j+1}\ldots]$, in which the ordering of the external legs is fixed. The full tree-level superamplitude is then obtained in the usual way by summing over all non-cyclic permutations of external legs and multiplying each partial amplitude with a single colour trace over the gauge group generators corresponding to each external leg in the order that they appear. See e.g. \cite{Mangano:1990by}. In the case of interest here, some simple structure to the non-zero colour-traces can be used to identify possible orderings of the massive and massless vector multiplets. 

As discussed in \cite{Craig:2011ws}, because of the bifundamental nature of the massive vector multiplets with respect to the unbroken gauge group factors, partial amplitudes must be of the form $\mathcal{A}_n[\mathcal{W}_{mi},G_i,G_i,\ldots\overline{\mathcal{W}}_{ij},G_j,G_j,\ldots\mathcal{W}_{jn},\ldots]$. Here, $G_i$ and $G_j$ are massless vectors of different unbroken gauge subgroups $SU(N_{i})$ and $SU(N_{j})$ respectively, while e.g. $\overline{\mathcal{W}}_{ij}$ has one fundamental $SU(N_{i})$ index and one antifundamental $SU(N_{j})$ index, so must be ordered to the left of a string of $G_j$ fields and to the right of a string of $G_i$ fields. The strings of massless vectors (of possibly zero length) can only terminate at a massive vector field with opposite index structure. Note that the overbar on the massive vectors merely distinguishes those with negative central charge (``anti-BPS'') from those with positive central charge. BPS and anti-BPS vectors need not alternately appear in the colour-ordered partial amplitudes for a general breaking pattern of the gauge group, but both must be present. In the subsequent discussion, we will not bother to distinguish between the vector multiplets belonging to different gauge subgroups, but will leave this implicit and fully encapsulated in the stripped colour trace.

Having established the colour-structure of the superamplitudes, we are now able to focus our attention on the more interesting kinematic structure of the superamplitudes with massive multiplets. The first feature to note is that all Coulomb branch superamplitudes $\mathcal{A}_n$ will be of homogeneous Grassmann degree $2n$ in our representation. This is a consequence of the $U(1)$ factor of the explicit $U(2)\leq USp(4)$ represented on the massive on-shell superspace. This subgroup is generated by the trace of the $\mathbf{m}^a_b$ generators in (\ref{MasslessR}) and (\ref{MassiveR}). As the vector bosons are $R$-invariant and the Grassmann variables carry a unit of charge under this generator, the massive superfield (\ref{LongMultNtwo}) must carry $2$ units of this $R$-charge. As the component amplitudes must conserve this charge, the $2$ units per leg in the superamplitude must be instead carried by accompanying Grassmann variables.

In this non-chiral superspace, the helicity-violating sectors of the massless superamplitudes appear as terms with Grassmann variables divided differently between $\eta^a$ and $\tilde{\eta}^{\dagger a}$ factors. This is clear from the contributions to the supercharges from the massless legs in (\ref{eqn:coulombscharges}), where $Q^{\dagger a+2}$ and $Q_{a}$ will not mix the sectors of definite helicity violation. However, both types of supercharges act on the massive Grassmann variables, so this structure is not respected by the massive legs. This is to be expected, because helicity is no longer a frame-independent property for massive particles. We will discuss how mass affects the sectors further below once we begin to compute higher leg amplitudes.

Finally, we adopt the convention that all particles are outgoing and that incoming states may be obtained by crossing outgoing legs. Under crossing, an outgoing leg of momentum $p$ is analytically continued to an incoming leg of momentum $-p$ and opposite central charge. The mass of the leg is unchanged, but a negative sign now accompanies its appearance in the Weyl equation and the spin sums (see \cite{HKT:2018a} for relevant identities in the conventions employed here). This is commented upon further below.

\section{Massive Super-BCFW Recursion}\label{sec:massiveBCFW}

\subsection{Massless Super-BCFW}

We will demonstrate below that supersymmetry fully determines the superamplitude with three external states. With more legs, supersymmetry is not enough and further properties of the $S$-matrix are required. To make progress in constructing higher-leg superamplitudes we will make use of BCFW recursion at tree level \cite{Britto:2004ap,Britto:2005fq}.
A BCFW shift on legs $i$ and $j$ consists, at the level of momenta, of finding a (complex) vector $r^\mu$ such that $p_i \cdot r = p_j \cdot r = r \cdot r = 0$, and shifting the two momenta to $p_i^\mu \rightarrow \hat{p}_i^\mu = p_i^\mu + z r^\mu$, $p_j^\mu \rightarrow \hat{p}_j^\mu = p_j^\mu - z r^\mu$, with $z$ a complex parameter. Note that this also necessitates shifting the polarisations of $i$ and $j$ as well, to maintain transversity. For massless legs, both of these deformations may be formulated simply at the level of spinors. An $[i,j\rangle$-shift is realised on the spinors as $\hat{\rs{i}}=\rs{i}+z\rs{j}$ and $\hat{\ra{j}}=\ra{j}-z\ra{i}$ (so the shift vector $r=-\rs{j}\la{i}$). A shift is called valid if the amplitude vanishes as $z \rightarrow \infty$. Cauchy's theorem then relates the value of the unshifted amplitude to a sum over complex poles of the shifted amplitude, which by tree-level unitarity occurs on on-shell factorization channels. 

The supersymmetric extension of on-shell recursion, known as super-BCFW \cite{Brandhuber:2008pf,ArkaniHamed:2008gz,Drummond:2008cr}, allows us to construct full superamplitudes recursively. It has been shown that any amplitude of pure Yang-Mills and matter containing a negative helicity gluon is on-shell constructible under a BCFW shift \cite{Cheung:2008dn}. For $\mathcal{N}=4$ at the origin of moduli space, the fact that the other states are related supersymmetrically to the negative helicity gluon suffices to show that all superamplitudes are constructible using a supersymmetric extension of BCFW \cite{ArkaniHamed:2008gz}. These arguments do not rely on the masslessness of the other legs of the superamplitude and consequently this shows that any Coulomb branch superamplitude which has two massless legs is on-shell constructible under a super-BCFW shift. 

The supersymmetrised BCFW-shift involves the standard BCFW shift described above supplemented with a shift in Grassmann variables to preserve the supercharge. For an $[i,j\rangle$-shift, the Grassmann variables are also shifted to $\hat{\eta}^A_i=\eta^A_i+z\eta^A_j$ in the chiral superspace. This may be derived by deducing the necessary shift in the supercharge $Q^{\dagger A}_i=\sqrt{2}\ra{i}\eta_i^A$ carried by leg $i$ resulting from demanding both that the total supercharge be conserved and that the SUSY algebra (\ref{FullSUSYalgebra}) be preserved (note that the derivatively represented $Q_{i,A}$ must also shift).

The standard super-BCFW shift may be converted into a form where it may be used in the non-chiral superspace. This can be obtained by half-Fourier transforming the shifted superamplitude in the chiral superspace. To implement a $[i,j\rangle$-supershift, the momentum shift is unchanged from that described above, while the Grassmann variables shift as $\hat{\eta}^a_i\rightarrow\eta_i^a+z\eta^a_j$ and $\hat{\tilde{\eta}}^{\dagger a}_{j}\rightarrow\tilde{\eta}^{\dagger a}_{j}-z\tilde{\eta}^{\dagger a}_{i}$. Constructibility continues to hold in this superspace, as the half-Fourier transform from the chiral superspace does not affect the large $z$ scaling of the superamplitude with shifted momentum.

\subsection{Massive BCFW}\label{MassiveBCFWNoQ}

While this standard super-BCFW shift is a powerful tool for constructing higher-leg Coulomb branch superamplitudes, it leaves open the question of constructing fully massive Coulomb branch superamplitudes. One path toward the on-shell construction of such superamplitudes is to formulate a supershift on massive legs. BCFW recursion for massive legs has been introduced in \cite{Badger:2005zh} and \cite{Schwinn:2007ee}. As in the massless case, the momenta shift as 
\begin{equation}
\hat{p}_i^\mu = p_i^\mu + z r^\mu, \qquad \hat{p}_j^\mu = p_j^\mu - z r^\mu,
\end{equation}
where $r$ has the same orthogonality properties as in the massless case. To construct the shift vector $r$, we find a little group frame for each particle where we can write $p_i$ and $p_j$ as linear combinations of the same two null vectors. Geometrically, these correspond to the two null vectors being coplanar with both massive momenta. Finding this little group frame requires solving
\begin{equation}
- \rs{i^2} \la{i^1} = \frac{\alpha_i}{m_j^2} \rs{j^1} \la{j^2}, \qquad - \rs{j^2} \la{j^1} = \frac{\alpha_j}{m_i^2} \rs{i^1} \la{i^2} \\
\end{equation}
to find $\alpha_i = \alpha_j \equiv \alpha$, where
\begin{gather}
\alpha = - p_i \cdot p_j + \sqrt{(p_i \cdot p_j)^2 - m_i^2 m_j^2} \\
p_i = \rs{i^1} \la{i^2} + \frac{\alpha}{m_j^2} \rs{j^1} \la{j^2},  \qquad p_j = \rs{j^1} \la{j^2} + \frac{\alpha}{m_i^2} \rs{i^1} \la{i^2}.
\end{gather}
Up to a single ambiguous phase, the spinors of each leg may be related in this special frame by
\begin{align}
\rs{i^1}=\frac{m_i}{\sqrt{\alpha}}\rs{j^2}\qquad
&\rs{i^2}=-\frac{\sqrt{\alpha}}{m_j}\rs{j^1}\nonumber\\
\ra{i^2}=-\frac{m_i}{\sqrt{\alpha}}\ra{j^1}\qquad&\ra{i^1}=\frac{\sqrt{\alpha}}{m_j}\ra{j^2}.\label{BCFWFrame}
\end{align}

\noindent In this special little group frame, it is clear that we may take 
\begin{equation}
r = \rs{i^1} \la{j^2} \text{ or } r = \rs{j^1} \la{i^2}\label{shift}
\end{equation}
and satisfy the orthogonality requirements $p_i \cdot r = p_j \cdot r = r \cdot r = 0$ \cite{Boels:2010mj}. It is clear that $r$ cannot be regarded merely as a function of the massive momenta $p_1$ and $p_2$, as it is determined by only a single helicity spinor associated to each. Its selection explicitly breaks little group invariance of the legs, as its existence relies on this preferred null vector decomposition.

The massive BCFW recursion may be illustrated on a simple example. Bhabha scattering in scalar QED is a constructible example, provided that, in the Lagrangian picture, there is a quartic scalar interaction with $-\frac{1}{2}e^2(\phi^*\phi)^2$ for electric charge $e$ (calling $\phi$ the scalar field) \cite{elvang2015scattering}. The validity of the shift may be verified by derivation from the Feynman rules, from which it can be shown that the shifted amplitude $A(\phi,\phi^*,\phi,\phi^*)\rightarrow 0$ as $z\rightarrow\infty$. This is not unexpected, as this amplitude is well-known to be constructible by BCFW recursion when the scalars are massless, provided that the shifted particles have the same charge. Unlike for spinning particles, massive scalars do not carry more degrees of freedom than massless scalars. When the massive legs are spinning, the validity of recursion is expected to be less general. The validity of massive BCFW for QCD amplitudes with massive quarks was discussed in \cite{Schwinn:2007ee}, which was spin-dependent. However, the case of massive scalars here does not introduce any substantial change.

Shifting the scalar $\phi$ legs $1$ and $3$, the amplitude is determined as a sum over two factorisation channels:
\begin{align}
A(\phi_1,\phi^*_2,\phi_3,\phi^*_4)=&\sum_{h=+,-}\hat{A}(\hat{\phi}_1,\phi^*_2,\gamma_{\hat{P}_{12}}^h)\frac{-1}{s}\hat{A}(\hat{\phi}_3,\phi^*_4,\gamma_{-\hat{P}_{12}}^{-h})\Big|_{z_*^{(1)}}\nonumber\\
&+\sum_{h=+,-}\hat{A}(\hat{\phi}_1,\phi^*_4,\gamma_{\hat{P}_{14}}^h)\frac{-1}{u}\hat{A}(\hat{\phi}_3,\phi^*_2,\gamma_{-\hat{P}_{14}}^{-h})\Big|_{z_*^{(2)}},\label{BhabhaBCFW}
\end{align}
where $\gamma$ is a photon and $\hat{P}_{12}=-\hat{p}_1-p_2$ and $\hat{P}_{14}=-\hat{p}_1-p_4$ are its (complex) momenta in each factorisation channel. The intermediate photon's helicity $h$ is summed over. The unshifted Mandelstam variables are $s=-(p_1+p_2)^2$, $u=-(p_1+p_4)^2$ and $t=4m^2-s-u$, for scalar mass $m$. The poles $z_*^{(i)}$ are determined by finding the values of the shift parameter $z$ on which the shifted momenta are aligned on a factorisation channel, but their identity will not be necessary here.

At this point, we review an exceptional feature which appears in the special case of three-leg amplitudes with two massive, equal-mass particles and one massless particle, such as $\hat{A}(\phi_1,\phi^*_2,\gamma_{3}^\pm)$. Introduced in  \cite{Arkani-Hamed:2017jhn}, an additional object that carries helicity weight of the massless particle exists that may be used as an amplitude building block:
\begin{equation}
x \equiv \frac{1}{m} \frac{\ls{q}p_2 \ra{3}}{\ds{q3}}, \label{eqn:xdef}
\end{equation}
where $3$ is the massless leg, $m$ is the mass of legs $1$ and $2$, and $\rs{q}$ is an arbitrary reference spinor defined so that $\ds{q3}\neq 0$. This special case arises because $p_2\cdot p_3=-\la{3}p_2\rs{3}=0$, implying that $p_2\rs{3}\propto \ra{3}$. The constant of proportionality is $x$ and carries helicity weight $1$ of leg $3$. It is independent of the reference spinor present in (\ref{eqn:xdef}). See \cite{HKT:2018a} for further details, conventions and identities.

The on-shell three-particle amplitudes in (\ref{BhabhaBCFW}) are
\begin{align}
A(\phi_1,\phi^*_2,\gamma_3^+)=\frac{em}{x}\\
A(\phi_1,\phi^*_2,\gamma_3^-)=emx.
\end{align}
Parity has been imposed. Denoting by $\hat{x}_{ij}$ the value of the $x$-factor at the shifted momentum in the three-leg amplitude with massive scalars $i$ and $j$, then for the purposes here
\begin{align}
\hat{x}_{12}=\frac{\ls{q}p_2\ra{\hat{P}_{12}}}{m\ds{q\hat{P}_{12}}}=\frac{m\da{\rho \hat{P}_{12}}}{\la{\rho}p_2\rs{\hat{P}_{12}}}\qquad\hat{x}_{34}=\frac{m\da{\rho (-\hat{P}_{12})}}{\la{\rho}p_4\rs{(-\hat{P}_{12})}}=\frac{\ls{q}p_4\ra{(-\hat{P}_{12})}}{m\ds{q(-\hat{P}_{12})}}\label{Bhabhax}
\end{align}
and similarly for $\hat{x}_{14}$ and $\hat{x}_{32}$. We leave implicit that these factors in (\ref{Bhabhax}) are to be evaluated on the pole $z=z_*^{(1)}$ while the others are on the $z=z_*^{(2)}$ pole. Here $\rs{q}$ and $\ra{\rho}$ are reference spinors not aligned with the spinors of the internal momentum $\hat{P}_{12}$. The $x$-factors are independent of the reference spinors. With these expressions, the Bhabha scattering amplitude is then
\begin{align}
A(\phi_1,\phi^*_2,\phi_3,\phi^*_4)&=\frac{-e^2m^2}{s}\left(\frac{\hat{x}_{12}}{\hat{x}_{34}}+\frac{\hat{x}_{34}}{\hat{x}_{12}}\right)+\frac{-e^2m^2}{u}\left(\frac{\hat{x}_{14}}{\hat{x}_{32}}+\frac{\hat{x}_{32}}{\hat{x}_{14}}\right)\nonumber\\
&=\frac{-e^2}{s}\left(\frac{\ls{q}p_2\ra{\hat{P}_{12}}\ls{(-\hat{P}_{12})}p_4\ra{\rho}}{\ds{q\hat{P}_{12}}\da{\rho (-\hat{P}_{12})}}+\frac{\ls{q}p_4\ra{(-\hat{P}_{12})}\ls{\hat{P}_{12}}p_2\ra{\rho}}{\ds{q(-\hat{P}_{12})}\da{\rho \hat{P}_{12}}}\right)\nonumber\\
&\qquad+(2\leftrightarrow 4)\nonumber\\
&=e^2(2p_2\cdot p_4)\left(\frac{1}{s}+\frac{1}{u}\right)=e^2(2m^2-t)\left(\frac{1}{s}+\frac{1}{u}\right).
\end{align}
See below in (\ref{MasslessAnaCont}) for spinor analytic continuation rules for negative momentum. Here, $\hat{P}_{ij}\cdot p_2=\hat{P}_{ij}\cdot p_4=0$ on either complex pole ($\{(i,j)=(1,2),(1,4)\}$) imply that these momenta anticommute as bispinors, while the Clifford algebra has been used in the step in which the reference spinors cancel out when the two terms for each channel are added together. This calculation is almost identical to the gluing argument of \cite{Arkani-Hamed:2017jhn}.

Unlike in (super)-Yang-Mills, BCFW here merely automates the construction of the amplitude from its two possible factorisation channels. However, unlike massless gauge theories, the second factorisation channel of the amplitude does not automatically emerge from the first. While the on-shell three-particle amplitudes contain ``non-local'' kinematic factors, these cancel in the sum over internal photon helicities, as explained in \cite{Arkani-Hamed:2017jhn}, along with the poles $z_*^{(i)}$. This happens regardless of the mass of the scalar legs. Foretelling further results below, at no point was the identity of the shift vector necessary in this computation. As the only source of little group violation, it cancelled-out in the end, being eliminated within each term in the BCFW expansion as part of the cancellation of the kinematic denominators upon each residue.

\subsection{Massive Super-BCFW}

Massless super-BCFW recursion has been established in $6$ dimensional \cite{Cheung:2009dc}, \cite{Dennen:2009vk} (and higher \cite{CaronHuot:2010rj}) super-Yang-Mills. In $6d$, the extra dimensions allow for extra directions in which the shift vector can point. As a result, the possible shift vectors are parameterised by an arbitrary variable in the massless $6d$ little group $SU(2)\times SU(2)$ (as it is effectively like a polarisation vector of one of the states).

The Coulomb branch of $4d$ SYM is equivalent to the low energy limit of the $6d$ theory after dimensional reduction on a torus (with fluxes providing the masses \cite{Scherk:1978ta}). The masses of the BPS states can be identified with the momenta in the compactified directions. The form of the supershift constructed here corresponds to the dimensional reduction of the $6d$ supershift defined in \cite{Dennen:2009vk}, having made the choice to align the six-dimensional shift vector along the four non-compact dimensions so that the $4d$ shift vector remains null. This reduces the possible $6d$ shifts to the two possibilities in $4d$ discussed above. It is presumably also possible to construct a super-shift for the Coulomb branch in which includes shifts to the masses. In the following, we will construct massive super-BCFW in $4d$ purely from consistency with the symmetry algebra and the non-supersymmetric shift constructed above.

In order to make the momentum shift supersymmetric, the supercharges of each leg must be deformed in order to preserve both the SUSY algebra (\ref{FullSUSYalgebra}) and the BPS constraint (\ref{BPS}). Demanding that the total supercharge still be conserved, the supercharges of the shifted legs become
\begin{align}
\frac{1}{\sqrt{2}}\hat{Q}_{i,a+2}=\frac{1}{\sqrt{2}}Q_{i,a+2}+\frac{z}{2}\Delta Q_{a+2}\qquad &\frac{1}{\sqrt{2}}\hat{Q}_{j,a+2}=\frac{1}{\sqrt{2}}Q_{j,a+2}-\frac{z}{2}\Delta Q_{a+2}\nonumber\\
\frac{1}{\sqrt{2}}\hat{Q}_i^{\dagger a}=\frac{1}{\sqrt{2}}Q_i^{\dagger a}+\frac{z}{2}\Delta Q^{\dagger a}\qquad &\frac{1}{\sqrt{2}}\hat{Q}_j^{\dagger a}=\frac{1}{\sqrt{2}}Q_j^{\dagger a}-\frac{z}{2}\Delta Q^{\dagger a}.
\end{align}
The derivatively represented supercharges in (\ref{eqn:coulombscharges}) also shift.

The shift spinors above may be expanded in a basis of Grassmann variables (or their derivatives) and spinors. The commutation relations and the BPS constraints may then be imposed in order to determine the coefficients. We will give the supercharge shift assuming that leg $i$ is BPS and leg $j$ is anti-BPS. All other particle/anti-particle configurations are also possible, but conservation of central charge implies that this configuration will at least always be available in any superamplitude. Explicitly choosing the special little group frame selected by the momentum shift and considering only $r=\rs{i^1}\la{j^2}$ for simplicity, the supercharges can be determined to shift as
\begin{align}
\Delta Q_{a+2} &= -\frac{2m_im_j}{\alpha+m_im_j}\rs{i^1}\left(\eta_{j1}^a+\frac{\sqrt{\alpha}}{m_i}\eta^a_{i2}\right)\label{Qshift}\\
\Delta Q^{\dagger a} &= -\frac{2m_im_j}{\alpha+m_im_j}\ra{j^2}\left(\eta_{i2}^a-\frac{\sqrt{\alpha}}{m_j}\eta^a_{j1}\right).\label{Qconjshift}
\end{align}
The supercharges shift in the spinor directions singled-out by the momentum shift vector. Note that these expressions may be converted into a form consisting of $r$ multiplying a little group invariant spinor expression. All little group violation may be contained to the shift vector $r$.

In contrast to the massless case, the BCFW shift implemented at the level of spinors and Grassmann variables has an ambiguity. This is because, while the shifted spinors of each leg are related through (\ref{BCFWFrame}), there is no analogue for the Grassmann variables. It is therefore possible to shift these by the Grassmann variables of the same leg, in addition to those of the other. This affects the numerical prefactor multiplying the spinor shift. Choosing the Grassmann variables to shift only by terms proportional to those of the opposite shifted leg, the supershift may be represented as:
\begin{alignat}{2}
\rs{\hat{i}^2}&=\rs{i^2}-z\frac{m_j\sqrt{\alpha}}{\alpha+m_im_j}\rs{i^1}\qquad &&
\la{\hat{i}^2}=\la{i^2}+z\la{j^2}\frac{m_im_j}{\alpha+m_im_j}\\
\rs{\hat{j}^1}&=\rs{j^1}-z\frac{m_im_j}{\alpha+m_im_j}\rs{i^1}\qquad &&
\la{\hat{j}^1}=\la{j^1}+z\la{j^2}\frac{m_i\sqrt{\alpha}}{\alpha+m_im_j}\\
\hat{\eta}^a_{i,1}&=\eta^a_{i,1}-z\frac{m_im_j}{\alpha+m_im_j}\eta^a_{j,1}\qquad && \hat{\eta}^a_{j,2}=\eta^a_{j,2}-z\frac{m_im_j}{\alpha+m_im_j}\eta^a_{i,2}
\end{alignat}
and the other components are unaffected. We are again only showing here the case for the momentum shift $r=\rs{i^1}\la{j^2}$.

The spinor-level shift of the massive legs may be re-expressed in a way that relates the little group violation directly to the momentum shift vector:
\begin{equation} 
\begin{split}
|\hat{i}_{I}]&=|i_{I}]+\frac{z}{2m_{i}}\rho|i_{I}\rangle-\frac{z}{2m_{i}m_{j}}p_{j}\rho|i_{I}] \\ 
\langle \hat{i}^{I}|&=\langle i^{I}|+\frac{z}{2m_{i}}[i^{I}|\rho+\frac{z}{2m_{i}m_{j}}\langle i^{I}|\rho p_{j} \\
|\hat{j}_{I}]&=|j_{I}]+\frac{z}{2m_{j}}\rho|j_{I}\rangle-\frac{z}{2m_{i}m_{j}}p_{i}\rho|j_{I}] \\
\langle \hat{j}^{I}|&=\langle j^{I}|+\frac{z}{2m_{j}}[j^{I}|\rho+\frac{z}{2m_{i}m_{j}}\langle j^{I}|\rho p_{i}.
\label{dimredshift}
\end{split}
\end{equation}
Here, $\rho \equiv\pm \left(m_im_j/\sqrt{(p_i\cdot p_j)^2-m_i^2m_j^2}\right) r$ ($(+)$ for $r=\rs{i^1}\la{j^2}$, $(-)$ for $r=\rs{j^1}\la{i^2}$), or equivalently $r_{\alpha \dot{\beta}}  = \frac{-1}{2 m_i m_j} \rho_{\alpha \dot{\alpha}} (p_i p_j - p_j p_i)_{\ \dot{\beta}}^{\dot{\alpha}}$. 
The corresponding shifts of the Grassmann variables are
\begin{equation}
\begin{split}
\hat{\eta}^{a}_{i,I}=\eta^{a}_{i,I}-\frac{z}{2m_{i}m_{j}}\Big([i_{I}|\rho|j^{J}\rangle - \langle i_{I}|\rho|j^{J}]\Big)\eta_{j,J}^{a} \\
\hat{\eta}^{a}_{j,J}=\eta^{a}_{j,J}-\frac{z}{2m_{i}m_{j}}\Big([j_{J}|\rho|i^{I}\rangle - \langle j_{J}|\rho|i^{I}]\Big)\eta_{i,I}^{a},
\end{split}\label{shiftgrass}
\end{equation}
where we have here again assumed that leg $i$ is BPS and leg $j$ is anti-BPS, although shifts with both legs of the same type are also possible and differ only in changes of signs both in (\ref{dimredshift}) and (\ref{shiftgrass}). In the ensuing calculations, we will not actually need any of these results beyond the existence of the momentum and supercharge shifts and their abstract properties. Rather, we merely state them here for completeness.

All Grassmann dependence of the superamplitudes arise in the form of the supercharges of each leg. Since the superfield legs are scalars, the supershift may be regarded entirely as a shift in momentum and supercharge by the null vector $r$ and chiral spinors presented above in (\ref{Qshift}) and (\ref{Qconjshift}). From this point of view, it is clear that the supershift vector and spinors do not obstruct the freedom in choosing little group decompositions of the momenta and supercharges of each unshifted leg. However, they provide a prefered null direction which singles out the little group frames in which both the shift vector and spinors have the especially simple forms (\ref{shift}), (\ref{Qshift}) and (\ref{Qconjshift}), leading to the apparent breaking of covariance in the spinor (\ref{dimredshift}) and Grassmann level shifts (\ref{shiftgrass}). The shift spuriously breaks the little groups of the shifted legs by providing a special direction in which the massive momenta may be decomposed. Use of super-BCFW will therefore preserve little group invariance of the recursed superamplitudes up to explicit appearances of the shift vector $r$. However, as this is the only source of the breaking and the superamplitude itself must be invariant, all appearances of the shift vector must ultimately cancel to leave a manifestly invariant expression. This is similar to the $6d$ perspective, where the cancellation of the shift also inevitably follows from the arbitrary and spurious choice of direction that must be made in choosing it.

This issue does not appear for $4d$ massless superamplitudes, where the bispinor form of the shift vector appears to manifestly break the $U(1)$ little group invariance. Because the residue on the complex pole scales as $z^{(i)}_\star\propto 1/r$ on each factorization channel $i$, the combination $z^{(i)}_\star r$ is a little group invariant function when $r$ is constructed out of massless helicity spinors. 

As mentioned in \cite{Arkani-Hamed:2017jhn}, for general massive amplitudes, the combination of different helicity states in the little group covariant formalism can obstruct on-shell constructibility, as not all helicity components have the correct large-$z$ behavior. However, as will be discussed in Section \ref{Sec:3leg}, supersymmetry forces the Coulomb branch three-leg superamplitudes to contain the precise ``nonlocality'' needed for them to combine to give the pole structure of the four-leg superamplitude. This first hint of simple factorization properties remarkably extends to all Coulomb branch superamplitudes, as it turns out that all such superamplitudes are on-shell constructible via massive super-BCFW. 

\subsection{Validity}\label{Validity}
While the underlying origin is likely a vestige of dual (super)conformal invariance remaining on the Coulomb branch, we here leave an exploration of this to future work and instead prove the shift validity by using soft limits to extend the known behavior at the origin of moduli space. The idea that Coulomb branch component amplitudes may be found from soft limits of massless amplitudes with scalar insertions was proposed in \cite{Craig:2011ws}, expanded upon in \cite{Kiermaier:2011cr} and proven in \cite{Elvang:2011ub}. The precise map is explained clearly around (4.3) of \cite{Kiermaier:2011cr}, but the details will not be necessary for us. All we rely on is the fact that the Coulomb branch component amplitudes may be written as a sum over amplitudes at the origin of moduli space.

We may utilize this relation to show that a massive super-BCFW shifted Coulomb branch superamplitude $\mathcal{A}\left[\lbrace\hat{\lambda}_{1I},\hat{\eta}^a_{1I}\rbrace,\lbrace\hat{\lambda}_{2I},\hat{\eta}^a_{2I}\rbrace,\dots\right]$ has the correct large $z$ scaling for a valid shift (where we are borrowing the notation of \cite{ArkaniHamed:2008gz} to highlight both the momentum spinors and Grassmann variables of each leg). The first step is to perform a $z$-independent supertranslation which sets $\hat{\eta}^a_{1I},\hat{\eta}^a_{2I} \rightarrow 0$: 
\begin{gather}
\mathcal{A}\left[\lbrace\hat{\lambda}_{1I},\hat{\eta}^a_{1I}\rbrace,\lbrace\hat{\lambda}_{2I},\hat{\eta}^a_{2I}\rbrace,\lbrace\lambda_{3I},\eta^a_{3I}\rbrace,\dots\right] = \mathcal{A}\left[\lbrace\hat{\lambda}_{1I},0\rbrace,\lbrace\hat{\lambda}_{2I},0\rbrace,\lbrace\lambda_{3I},\eta^a_{3I} \mp \da{3_I \zeta} - \ds{3_I \zeta}\rbrace\dots\right] \label{eqn:massivetrans}\\
\ra{\zeta} = \frac{1}{s_{12}}\left( p_2 \rs{1^I} \eta_{1I}^a - p_1 \rs{2^I} \eta^a_{2I} - m_2 \ra{1^I} \eta^a_{1I} + m_1 \ra{2^I} \eta^a_{2I}\right) \\
\rs{\zeta} = \frac{1}{s_{12}}\left( - p_2 \ra{1^I} \eta_{1I}^a - p_1 \ra{2^I} \eta^a_{2I} + m_2 \rs{1^I} \eta^a_{1I} + m_1 \rs{2^I} \eta^a_{2I}\right), 
\end{gather}
where we have assumed that leg $1$ is BPS and leg $2$ is anti-BPS in our explicit solutions for $\ra{\zeta},\rs{\zeta}$, but an analogous procedure may be done for any two shifted legs with either sign central charge. This is a supersymmetry transformation which relates all component amplitudes to those with two of the lowest-weight states, which are here the scalars $\phi$. The existence of such a transformation that sets the shifted Grassmann variables to zero while not reintroducing $z$ into the other Grassmann variables was first pointed out in the massless case in \cite{ArkaniHamed:2008gz}. 

Each massive component of (\ref{eqn:massivetrans}) is given by \cite{Kiermaier:2011cr} as a soft limit of a sum of massless component amplitudes with scalar insertions. Importantly, all the components of the translated superamplitude (\ref{eqn:massivetrans}) have two shifted lowest-weight scalars $\phi$, so for any massive component amplitude the sum will be over massless amplitudes with two shifted lowest-weight scalars $S_{34}$. Schematically, for any component of (\ref{eqn:massivetrans}) we have 
\begin{equation}\label{eqn:softlim}
A\left[\hat{\phi}_1,\hat{\phi}_2,\dots\right] \sim \lim \sum A\left[\widehat{S}_{34},\varphi_{\text{vev}},\dots,\varphi_{\text{vev}},\widehat{S}_{34},\varphi_{\text{vev}},\dots\right],
\end{equation}
where the left side is a Coulomb branch amplitude and the right side is a sum over amplitudes at the origin of moduli space with insertions of scalars $\varphi_{\text{vev}}=-\frac{1}{2}(S_{13} - S_{24})=-\Re (S_{13})$, which are the massless scalar degrees of freedom which gain a vev on the Coulomb branch. These fields are taken soft by the limit. Each massless amplitude on the right side of (\ref{eqn:softlim}) is obviously a component of some massless superamplitude 
\begin{equation}\label{eqn:masslesstrans}
\mathcal{A}\left[\lbrace\hat{\lambda}_1,0\rbrace,\lbrace\lambda_{\text{vev}},\eta_{\text{vev}},\tilde{\eta}^\dagger_{\text{vev}}\rbrace,\dots,\lbrace\lambda_{\text{vev}},\eta_{\text{vev}},\tilde{\eta}^\dagger_{\text{vev}}\rbrace,\lbrace\hat{\lambda}_2,0\rbrace,\lbrace\lambda_{\text{vev}},\eta_{\text{vev}},\tilde{\eta}^\dagger_{\text{vev}}\rbrace\dots\right]
\end{equation}
in non-chiral superspace, where the Grassmann variables of the two shifted lines have been set to zero.

It was shown in \cite{ArkaniHamed:2008gz} that all massless $\mathcal{N}=4$ superamplitudes scale as $1/z$ in chiral superspace and, as mentioned above, the half-Fourier transform to non-chiral superspace does not modify the scaling. For any such superamplitude we may then perform the massless version of the supertranslation above to rid the superamplitude of the shifted Grassmann variables and bring it to the form of (\ref{eqn:masslesstrans}), where the shifted legs are lowest-weight scalars. This removes any factors of $z$ from the Grassmann monomials, so the superamplitude scaling immediately implies that the individual components of (\ref{eqn:masslesstrans}) must vanish as $1/z$. Then, from (\ref{eqn:softlim}), the massive components, as sums of amplitudes scaling as $1/z$, must also scale as $1/z$. The translated massive superamplitude in (\ref{eqn:massivetrans}) also has no $z$ dependence in its Grassmann variables, and so we may argue in reverse and upgrade the $1/z$ scaling of the component amplitudes to that of the full superamplitude. Thus the massive super-BCFW shifted superamplitude vanishes at infinity and therefore this is a valid shift.

This proves the validity of massive super-BCFW shifts of Coulomb branch superamplitudes. In concert with the aforementioned validity of super-BCFW when massless legs are shifted, this shows that all Coulomb branch superamplitudes are super-BCFW constructible.

\section{Scattering Amplitudes on the $\mathcal{N}=4$ Coulomb Branch}\label{MassiveAmplitudes}

The study of amplitudes at the origin of moduli space has revealed surprising structures and remarkable simplicity. 
The question of how much of this survives with massive states is not only of intrinsic interest, but also has use in understanding the loop-level properties of the massless theory. A first attempt to construct massive amplitudes and trace the way that the massless amplitudes are deformed by Higgsing was made in \cite{Craig:2011ws}. They used a superspace representation analogous to that traditionally used at the origin of the moduli space, in which the full $R$-symmetry is manifest (although the little group is not). Of particular note for the discussion here is that they were able to deduce that the superamplitudes could be decomposed into distinct `band' structures interrelated by SWIs, analogous to the usual sectors classified by degree of helicity violation, as well as find explicit expressions for the simplest cases of these. The use of soft limits discussed above was also proposed, which was then expanded upon in \cite{Kiermaier:2011cr} to reconstruct tree-level amplitudes as a series expansion in mass.

After hints arising in loop computations (see e.g. \cite{Bern:2006ew,Drummond:2006rz}), dual conformal symmetry was discovered in massless gluon amplitudes at strong coupling through holographic computations in \cite{Alday:2007hr}, where it was shown that this was the conformal symmetry associated with Wilson loops $T$-dual to the amplitude. As mentioned above, this symmetry has also been discovered in the planar amplitudes at weak coupling, thereby suggesting some non-perturbative property of the theory that may be accessible through analytic techniques. At leading order, dual conformal symmetry is enhanced to superconformal and combines with spacetime superconformality into a Yangian symmetry (see e.g. \cite{Beisert:2010jr} for review of integrability in $\mathcal{N}=4$ SYM). The breaking of dual conformality at loop level by IR divergences is understood from the Wilson loop duality \cite{Drummond:2007au} and has been used to fully determine the amplitudes with fewer than six legs \cite{Bern:2005iz}. However, the extent of the usefulness and survival of the enhancements in scattering amplitudes at loop-level is still under investigation \cite{Korchemsky:2009hm,Bargheer:2009qu,Sever:2009aa,CaronHuot:2011kk,Kanning:2018moy}, although some of the progress has made use of this at the level of the loop integrands where infrared divergences can be sidestepped.

Loop-level investigations into dual conformal symmetry led to the suggestion of using Higgsing as a way of regulating IR divergences in loop amplitudes between massless particles \cite{Alday:2009zm}. This was subsequently used in \cite{Henn:2010bk,Henn:2010ir} to constrain the form of loop integrands (an amusing application of this to computing the hydrogen spectrum in $\mathcal{N}=4$ SYM was shown in \cite{Caron-Huot:2014gia}). The resulting prediction of this symmetry that $1$-loop amplitudes do not involve triangle integrals (like their massless counterparts \cite{ArkaniHamed:2012nw}) was verified in \cite{Boels:2010mj}, where a massive on-shell superspace was also set-up. 

Following \cite{Huang:2011um}, it was attempted in \cite{Plefka:2014fta} to obtain massive amplitudes in $\mathcal{N}=4$ SYM in $4d$ by dimensional reduction from amplitudes in $6d, \ \mathcal{N}=(1,1)$ SYM. The $6d$ SYM amplitudes also feature dual conformal symmetry at tree-level (as well as at loop-level integrands), despite not being conformal themselves \cite{Dennen:2010dh} (this was also observed in $10d$ $\mathcal{N}=1$ SYM, which also reduces to this $6d$ theory \cite{CaronHuot:2010rj}). The realisation of this symmetry and the way that it is inherited by the massive $4d$ amplitudes, its possible relation to the Yangian and its usefulness in providing a guiding structure for determining the superamplitudes were discussed in \cite{Dennen:2010dh} and \cite{Plefka:2014fta}. The former used reduction of the $6d$ dual conformal symmetry to establish that the massive $\mathcal{N}=4$ tree amplitudes and loop integrands were dual conformal invariant, while some progress was made in the latter using this, as well as $6d$ super-BCFW recursion \cite{Dennen:2009vk}, to build some superamplitudes at low numbers of legs with manifest dual conformal symmetry (tests at loop-level were made in \cite{Bern:2010qa}). The symmetry algebra has been more recently discussed in \cite{Bering:2018yyp}. However, a general procedure for efficiently computing higher leg amplitudes is still left outstanding. As stated in the introduction, \cite{Cachazo:2018} (and, more recently, \cite{Geyer:2018xgb}) introduce CHY formulae for all $6d$ $\mathcal{N}=(1,1)$ massless amplitudes which may reduced to give a general formula for all $4d$ massive $\mathcal{N}=4$ tree amplitudes in a CHY form. It remains to be seen exactly how these special symmetries affect the structure of the massive amplitudes and can be used to explicitly construct them. 
The first step in such an investigation is to calculate and dissect some tree amplitudes in a presentable way. Once the patterns are identified, they can be used to guide the development of systematic computational techniques. We do this in the hope that it will ultimately help in grappling with the way in which the aspects of spin, supersymmetry and dual conformal symmetry interplay. 

The first steps toward elucidating the special symmetries of the Coulomb branch amplitudes is to compute the simplest examples and search for the patterns. It is this goal that we initiate in the remainder of this section.

\subsection{Special Massive Kinematics and Three Particle Superamplitudes}\label{Sec:3leg}

\subsubsection{Special BPS Kinematics}
Similarly to their massless counterparts, on-shell three-particle amplitudes of massive BPS vector multiplets exhibit special kinematical properties. Without loss of generality, we will consider the superamplitude $\mathcal{A}_3[\mathcal{W},\overline{\mathcal{W}},\mathcal{W}]$ with two BPS and one anti-BPS states. Conservation of the central charge implies that $m_2=m_1+m_3$. This configuration of masses yields precisely a massive analogue of the special $3$-particle kinematics of massless $3$-leg amplitudes, because restricting the momenta to be real implies that they are parallel.

The special kinematic features of the $6d$ three-particle amplitudes have been described in \cite{Cheung:2009dc}. The $4d$ BPS particles have analogous properties. We will introduce these following the presentation in \cite{Cheung:2009dc} before giving a more geometric account further below.

It is simple to show that $\ds{i^Ij^J}\pm\da{i^Ij^J}$ has vanishing determinant as a matrix in little group indices for any pair of legs $i$ and $j$, where the $(-)$ is to be chosen if $i$ and $j$ have central charges of opposite sign and $(+)$ is chosen if they are the same. This implies the factorisation
\begin{align}
\ds{i^Ij^J}\pm\da{i^Ij^J}=u_i^Iv_j^J
\end{align} 
for some pure (complexified) $SU(2)$ spinors $u_i$ and $v_i$. It follows from these equations and the spin sums that $u_{i,I}\ls{i^I}\propto v_{i,I}\ls{i^I}\propto u_{j,I}\ls{j^I}$ for each $i$ and $j$ and likewise $u_{i,I}\la{i^I}\propto v_{i,I}\la{i^I}\propto u_{j,I}\la{j^I}$ and that $u_i^I\propto v_i^I$ for each leg - it is only distinct $GL(1)$ rescaling redundancies that distinguishes $u_i$ from $v_i$. These $GL(1)$ rescaling freedoms represent the complexification of the $U(1)$ ``tiny groups'' of each pair of particles \cite{Boels:2012ie}. This is the subgroup of Lorentz transformations that stabilises a pair of massive momenta. However, we find that this doubling is practically unnecessary and fix the scales so that $v_i=u_i$ for each $i$. The following identities then hold:
\begin{align}
\ds{1^I2^J}-\da{1^I2^J}&=u_1^Iu_2^J\nonumber\\
\ds{2^J3^K}-\da{2^J3^K}&=u_2^Ju_3^K\label{MassiveFactorize}\\
\ds{3^K1^I}+\da{3^K1^I}&=u_3^Ku_1^I\nonumber
\end{align}
and they imply 
\begin{align}
u_{1,I}\la{1^I}&=u_{2,J}\la{2^J}=u_{3,K}\la{3^K}\equiv \la{u}\nonumber\\
u_{1,I}\ls{1^I}&=-u_{2,J}\ls{2^J}=u_{3,K}\ls{3^K}\equiv \ls{u}.\label{M3PSK}
\end{align}

The spinors in a general little group frame may therefore be decomposed into components in this special frame in which the null vector decomposition ``aligns'' in this complexified way. The little group spinors may be decomposed into a magnitude and direction as $u_i^I=|u_i|\hat{u}_i^I$, where $\hat{u}_i^I$ is a unit $SU(2)$ spinor. To construct a little group basis including $u_{i,I}$, define
\begin{align}\label{protowbasis}
\hat{w}_{i,I}=\hat{u}^\dagger_{i,I}+\omega_i\hat{u}_{i,I},
\end{align}
as linearly independent spinor, where $\omega_i\in\mathbb{C}$ is free (so $\hat{w}_{i}$ need not be a unit spinor). A little group spinor basis may be completed with
\begin{align}
w_{i,I}=\frac{1}{|u_i|}\hat{w}_i. \label{wbasis}
\end{align}
This is effectively a dual spinor and satisfies $w_{i}^Iu_{i,I}=1$. This condition is necessary for the momenta to be on-shell (and the overall sign is fixed by requiring that the momenta be future-pointing in the real limit). The momenta may then be decomposed as
\begin{align}\label{MomentumDecomp}
p_i=w_{i,I}\rs{i^I}\la{u}\mp \rs{u}\la{i^J}w_{i,J}
\end{align}
(the $(+)$ is for BPS, $(-)$ for anti-BPS). The real momentum limit corresponds to $\omega_i\rightarrow 0$, but for complex momenta, $\omega_i$ is an undetermined residual redundancy. In this latter case, the spinors in each term in the decomposition are not complex conjugates and the little group is complexified from $SU(2)$ to $SL(2,\mathbb{C})$. 

The resemblance of (\ref{MomentumDecomp}) with the $3$-particle massless special kinematics is clear. As will be shown further below, in the high energy limit, the null vectors in one of these sets all shrink to zero, recovering the usual special $3$-particle kinematics for massless particles.

Calling the spinors $\hat{w}_{i,I}\ra{i^I}=\ra{i^w}$ (and similarly for the left-handed spinors), then
\begin{align}
\ds{i^Ij^J}\pm\da{i^Ij^J}=\hat{u}_i^I\hat{u}_j^J\left(\ds{i^{w}j^{w}}\pm\da{i^{w}j^{w}}\right),\label{SpinorMatrix}
\end{align}
The $\hat{u}_i$ give the direction in which the little group matrix on the left hand side of (\ref{SpinorMatrix}) has its only non-zero entry given by the accompanying factor. The combination
\begin{align}\label{non0fact}
|u_i||u_j|=\ds{i^{w}j^{w}}\pm\da{i^{w}j^{w}}
\end{align}
is here the massive analogue of the spinor bilinears of massless particles like $\da{ij}$ and $\ds{ij}$, only one of which is non-zero, as determined by the configuration of special $3$-particle massless kinematics. We will see below that the special massive kinematics will imply that the amplitudes will be functions of this combination of bilinears, along with accompanying little group tensor factors that encode polarisation information. However, first note that these relations may be inverted to give
\begin{align}
|u_1|=\sqrt{\frac{\left(\ds{1^{w}2^{w}}-\da{1^{w}2^{w}}\right)\left(\ds{3^{w}1^{w}}+\da{3^{w}1^{w}}\right)}{\left(\ds{2^{w}3^{w}}-\da{2^{w}3^{w}}\right)}}\label{scale}
\end{align}
and similarly for the others. Note that because the $u_i$ spinors carry the scale in (\ref{scale}), they contain more information than merely a preferred little group decomposition in which the spinors of each leg align. This discussion is also entirely independent of the choice of $\omega_i$ variables in the definition of $w_i$ frame spinors (\ref{protowbasis}).

As is clear in (\ref{MomentumDecomp}), for a single leg, the little group basis choice $\{u_i,w_i\}$ still has a remaining $GL(1,\mathbb{C})$ little group freedom under which $u_i$ and $w_i$ rescale oppositely. The choice of scales given by the equations (\ref{M3PSK}) reduces this to a single $GL(1,\mathbb{C})$ for all three legs. In the special case of real momenta, (\ref{M3PSK}) also leaves the overall direction free. However, this is fixed by the complex deformation.

Geometrically, a massive momentum vector may be decomposed into a sum of two future-directed, light-like vectors. The $SU(2)$ little group invariance represents the manifold of all such decompositions. Each null vector rotates about the massive vector under the Wigner rotations, but their sum remains unchanged. When the three leg momenta are real and parallel, each can be decomposed into a linear combination of the same two real null vectors. However, when complexified, the massive leg momenta need no longer be proportional and the requirement that their null vector decompositions coincide becomes a more stringent constraint. Here, ``coincide'' means that they have vanishing dot products (as is clear for e.g. the first terms in (\ref{MomentumDecomp}) for each $i$). However, in this complexified context, null vectors with vanishing dot product need no longer be linearly dependent (proportional). This is the complexification of the notion of massive momenta being parallel. The spinors constituting each null vector are also no longer complex conjugates. As a result, the little group is also complexified from $SU(2)$ to $SL(2,\mathbb{C})$.

The existence of the preferred spinor direction is analogous to that provided by the massless particle in the general $3$-leg amplitude with one massless leg and two massive legs of equal mass, as classified in \cite{Arkani-Hamed:2017jhn}. It is therefore analogously possible to construct general $3$-particle amplitudes between massive particles obeying this mass constraint by expanding the polarisation-stripped amplitude tensor in a basis spanned by tensor products of $\rs{u}_\alpha$ and $\epsilon_{\alpha\beta}$. However, by Lorentz invariance, the amplitude tensor must be of even total rank, so factors of $\rs{u}_\alpha$ may always be  paired and eliminated using (\ref{MassiveFactorize}) and (\ref{M3PSK}). Doing so will always leave (after applying the spin sums) terms proportional to $\epsilon_{\alpha\beta}$ and $(p_1p_2)_{\alpha\beta}$, which are precisely the building-blocks used for the general $3$-leg amplitude of three massive particles proposed in \cite{Arkani-Hamed:2017jhn}. Thus these special kinematics provide no new constraints on possible Lorentz structures in amplitudes nor any new features beyond the general case.

However, the special case in which one leg is massless and the other two have equal mass can be regarded as a limiting case. Taking $m_3\rightarrow 0$ and $m_2\rightarrow m_1=m$, then in the helicity basis for the little group frame,
\begin{align}
u_{3+}\rightarrow\pm\sqrt{m x}\qquad u_{3-}\rightarrow\pm\sqrt{m/x}\label{u3limits}
\end{align}
(the sign choice in each limit is to be the same). These components of the frame spinor produce the helicity-weight-carrying scalar units $x=u_{3+}/u_{3-}$, introduced in \cite{Arkani-Hamed:2017jhn}. The remaining $u_1$ and $u_2$ spinors can still be used as building-blocks, but can be related to $x$ and $\ra{3}$ and $\rs{3}$ through (\ref{MassiveFactorize}). Explicitly,
\begin{align}
u_1^I=\mp\sqrt{\frac{x}{m}}\ds{1^I3}\qquad u_2^J=\pm\sqrt{\frac{x}{m}}\ds{2^J3}.\label{otheru3limits}
\end{align}

Nevertheless, the case of a massless leg with two massive legs of equal mass is distinguished from other cases obeying the mass selection rule in that it does not have a non-trivial, real momentum, collinear limit. As a result, $x$ is a purely complex momentum object with no analogue in amplitudes of other mass configurations. The significance of this was observed recently by \cite{Caron-Huot:2018ape}, where it was shown that amplitudes of magnetic monopoles factorise differently on each of the two possible complex momentum configurations corresponding to the same factorisation channel, which was interpreted as signifying the presence of a Dirac string. In contrast, the Bhabha scattering calculation presented in Section \ref{MassiveBCFWNoQ} illustrates the simplest way in which the $x$ factors across a factorisation channel can be combined that does not depend upon the complex momentum configuration chosen, see discussion in \cite{Caron-Huot:2018ape}.

\subsubsection{Three-Particle Superamplitude}

To begin with, we present the $3$-particle superamplitude for massless legs in non-chiral superspace, which is
\begin{align}
\mathcal{A}_3[G_1,G_2,G_3]=&\frac{1}{\da{12}\da{23}\da{31}}\delta^{(4)}(Q^\dagger)\prod_a(\da{12}\tilde{\eta}^{\dagger a}_3+\da{23}\tilde{\eta}^{\dagger a}_1+\da{31}\tilde{\eta}^{\dagger a}_2)\nonumber\\
&\quad+\frac{1}{\ds{12}\ds{23}\ds{31}}\delta^{(4)}(Q)\prod_a(\ds{12}\eta^a_3+\ds{23}\eta^a_1+\ds{31}\eta^a_2).\label{3massless}
\end{align}
The first term is the MHV sector and the second is the anti-MHV ($\overline{\text{MHV}}$) sector. Each term is only non-zero for distinct special massless kinematical configurations. This may be obtained from the well-known chiral form by the half-Fourier transform. We henceforth choose to absorb the annoying factor of $\sqrt{2}$ in the supercharges in (\ref{eqn:coulombscharges}) into the definition of the coupling so that it is implicitly to be omitted in all appearances of the delta functions $\delta^{(4)}(Q)$ and $\delta^{(4)}(Q^\dagger)$.

We next turn to deriving the superamplitude for massive legs. Usually, supersymmetry invariance of an amplitude immediately implies that $\mathcal{A}_n\propto\delta^{(4)}(Q^{\dagger a})\delta^{(4)}(Q_{a+2})$. However, as will be shown below, the special kinematics here implies that $2$ pairs of supercharges of each chirality degenerate, leaving only $6$ independent (if the momenta were restricted to be real, then all $4$ pairs would be related). This occurs as a result of the special spinor direction given by the $u_i^I$. It is simple to show that $\da{uQ^{\dagger a}}=-\ds{uQ_{a+2}}=\sum_im_iu_i^I\eta_{iI}^a$. In this case, the superamplitude may be deduced from the little group scaling of the external legs (which are invariant in this coherent state basis) and invariance under the independent supersymmetries. Building a supersymmetry invariant involves introducing a new reference spinor $\ra{q}\not\propto \ra{u}$, effectively to decompose the supercharges into the shared components that are parallel to $\ra{u}$ and the remaining independent components. Projected onto $\ra{u}$ and $\ra{q}$, the delta functions may be factorised as $\delta^{(4)}(Q^{\dagger a})=\frac{1}{\da{qu}^2}\delta^{(2)}(\da{q Q^{\dagger a}})\delta^{(2)}(\da{u Q^{\dagger a}})$ and $\delta^{(4)}(Q_{a+2})=\frac{1}{m_1^4\da{q u}^2}\delta^{(2)}(\la{q}p_1\rs{Q_{a+2}})\delta^{(2)}(\la{u}p_1\rs{Q_{a+2}})$, where both expressions are independent of $\ra{q}$. Up to a multiplicative prefactor, the supersymmetry invariant may be obtained by dropping the repeated factor in both $\delta^{(4)}(Q^{\dagger a})$ and $\delta^{(4)}(Q_{a+2})$. This is easily verified as being annihilated by all of the supercharges. To determine the numerical prefactor, we demand that the result match onto (\ref{3massless}) in the limit of massless legs. The superamplitude is thus determined to be
\begin{align}
\mathcal{A}_3[\mathcal{W}_1,\overline{\mathcal{W}}_2,\mathcal{W}_3]&=\frac{1}{m_1^2\la{q}p_1p_3\ra{q}}\delta^{(4)}\left(Q^{\dagger a}\right)\delta^{(2)}\left(\la{q}p_1\rs{Q_{a+2}}\right)\nonumber\\
&=\frac{1}{\la{q}p_1p_3\ra{q}}\delta^{(4)}\left(Q_{a+2}\right)\delta^{(2)}\left(\da{qQ^{\dagger a}}\right).
\label{3massive}
\end{align}
The superamplitude has been expressed in a form in which the auxiliary spinors $u_i$ do not appear explicitly, although they still constrain the reference spinor $\ra{q}$ to satisfy $\da{uq}\neq 0$. While $\delta^{(4)}(Q^{\dagger a})=\frac{1}{\da{qu}^2}\delta^{(2)}(\da{uQ^{\dagger a}})\delta^{(2)}(\da{qQ^{\dagger a}})$ is clearly independent of the reference spinor, this remains true of $\frac{1}{\da{qu}^2}\delta^{(2)}(\la{q}p_1\rs{Q_{a+2}})$ up to terms that vanish when multiplied by the other delta functions. It is therefore justified to the drop of the factor of $\delta^{(2)}(\la{u}p_1\rs{Q_{a+2}})$ in $\delta^{(4)}(Q)$ to obtain the SUSY invariant in the first form in (\ref{3massive}) (and a similar argument applies to the second). The reference spinor itself is unnecessary for the component amplitudes and may be eliminated after these are extracted. However, it is needed to squash them all into the superamplitude in this way. A similar representation of the massless three particle superamplitude in $6d$ was found in \cite{Boels:2012ie}, which presumably reduces to the expression above upon dimensional reduction.

Also of note is that this superamplitude combines terms that belong to distinct supersymmetric sectors ($\text{MHV}$ and $\overline{\text{MHV}}$ in the massless limit) into a single Grassmann polynomial. We will return to this point and see the combination of sectors even more explicitly in Section \ref{sec:5ptCoulomb}.

The most remarkable feature of the massive $3$-leg superamplitude (\ref{3massive}) is the kinematic factor in the denominator. This vanishes in the collinear limit - the one situation in which the momenta can be both real and on-shell. This factor is reminiscent of the Parke-Taylor factors of the exact, massless Yang-Mills $3$-leg amplitude e.g. $\sim\frac{\da{12}^4}{\da{12}\da{23}\da{31}}$, as well as its supersymmetrised counterpart $\frac{\delta(Q)}{\da{12}\da{23}\da{31}}$. For these theories, when glued into a $4$-leg amplitude on a factorisation channel, the factors in the denominator combine to produce the pole representing the other factorisation channel of the amplitude, as arranged for automatically by BCFW recursion \cite{Benincasa:2007xk}, \cite{Arkani-Hamed:2017jhn}. However, in the massive case here, the kinematic factor is neither present nor necessary in any of the component amplitudes. Instead, its appearance is orchestrated as a consequence of the maximal supersymmetry. Its presence likewise suggests that the Coulomb branch superamplitudes share in the special constructibility properties of their massless counterparts, as confirmed by the existence of super-BCFW. This will be explored further below.

An alternative representation of the three particle superamplitude also exists that more directly utilises the special kinematical properties of the BPS states. In the special frame selected by $\{u_i,w_i\}$, the multiplicative supercharges may be decomposed as
\begin{align}
Q=\rs{u}\sum_i\eta_{iw}-\sum_i\pm\frac{1}{|u_i|}\rs{i^{w}}\eta_{iu}\qquad Q^\dagger =-\ra{u}\sum_i\eta_{iw}+\sum_i\frac{1}{|u_i|}\ra{i^{w}}\eta_{iu},\label{superchargesaligned}
\end{align}
calling Grassmann variables $\eta_{iu}=u^I_i\eta_{i,I}$ and $\eta_{iw}=w^{I}_i\eta_{i,I}$ (not $\hat{w}_i$ as used in the definition of $\ra{i^w}$). Now, partially solving the supercharge conservation constraints $\ds{u Q}=0$ and $\da{u Q^\dagger}=0$ implies that $\eta_{iu}=\pm\eta_{ju}$ for all legs $i$ and $j$ (where the $(+)$ applies if the central charges of $i$ and $j$ are the same and $(-)$ if they are opposite). This consequently implies that, on the support of this solution, the supercharges are parallel to the special frame spinor directions e.g. $Q\sim\rs{u}\left(\sum_i\eta_{iw}-C\eta_{1u}\right)$. The constant $C$ may be determined by introducing the reference spinor $\rs{q}$ satisfying $\ds{qu}\neq 0$ (any of the $\rs{i^w}$ would be possible choices):
\begin{align}
C=\frac{1}{\ds{qu}}\sum_i\frac{1}{(u_i)}\ds{qi^w}.\label{ExtraConstant}
\end{align}
An alternative representation of the supersymmetric delta function may therefore be deduced by combining each of the three distinct Grassmann terms in (\ref{superchargesaligned}) above into a single product
\begin{align}
\mathcal{A}_3[\mathcal{W}_1,\overline{\mathcal{W}}_2,\mathcal{W}_3]&=\prod_a\left(\left(\sum_i\eta_{iw}^a\right)\left(\eta_{1u}^a\eta_{2u}^a+\eta_{2u}^a\eta_{3u}^a-\eta_{3u}^a\eta_{1u}^a\right)-C\eta^a_{1u}\eta^a_{2u}\eta^a_{3u}\right).\label{3particle2UnFixed}
\end{align}

Note that, thus far, every expression involving a decomposition into this special little group frame is independent of the choice of $\omega_i$ in (\ref{wbasis}). These parameters remain free. Further simplification may be achieved by partially fixing the $\omega_i$ parameters to set $C=0$, or equivalently
\begin{align}
\sum_i\frac{1}{|u_i|}\rs{i^w}=0\qquad\sum_i\pm\frac{1}{|u_i|}\ra{i^w}=0
\end{align}
(these two equations are equivalent). On the support of each other's delta functions, the supercharges then reduce to the first terms in (\ref{superchargesaligned}). The superamplitude simplifies to 
\begin{align}
\mathcal{A}_3[\mathcal{W}_1,\overline{\mathcal{W}}_2,\mathcal{W}_3]&=\prod_a\left(\sum_i\eta_{iw}^a\right)\left(\eta_{1u}^a\eta_{2u}^a+\eta_{2u}^a\eta_{3u}^a-\eta_{3u}^a\eta_{1u}^a\right).\label{3particle2}
\end{align}
This is analogous to the form commonly presented in $6d$ \cite{Dennen:2010dh}.

The massive amplitudes are built out of these combinations of bilinears in (\ref{non0fact}). In (\ref{3particle2}), these are split apart into their `square roots' $|u_i|$. In extracting a component amplitude, four factors of $u_i$ and two of their duals $w_i=\frac{1}{|u_i|}\hat{w}_i$ are produced. These combine into spinor bilinears through (\ref{non0fact}). This demonstrates how the frame spinors $u_i^I$ can be used as alternative building blocks with which to construct the three particle amplitudes. 

To illustrate this more explicitly, the three massive vector component amplitude may be extracted from (\ref{3particle2}) to give
\begin{align}
A[W_1^{I_1I_2},\overline{W}^{J_1J_2}_2,W^{K_1K_2}_3]=\prod_{i=1}^2\left(\frac{|u_2||u_3|}{|u_1|}\hat{w}_1^{I_i}\hat{u}^{J_i}_2\hat{u}^{K_i}_3-\frac{|u_3||u_1|}{|u_2|}\hat{w}_2^{J_i}\hat{u}^{K_i}_3\hat{u}^{I_i}_1+\frac{|u_1||u_2|}{|u_3|}\hat{w}_3^{K_i}\hat{u}^{I_i}_1\hat{u}^{J_i}_2\right).\label{3massivevecComp}
\end{align}
The little group indices are implicitly to be symmetrised over (we will assume this in all subsequent expressions where they arise as indexing polarisation states of external legs). The diagonal terms in the product have the form e.g. 
\begin{align}
\prod_i\left(\frac{|u_2||u_3|}{|u_1|}\hat{w}_1^{I_i}\hat{u}^{J_i}_2\hat{u}^{K_i}_3\right)=\frac{\left(\ds{2^{w}3^{w}}-\da{2^{w}3^{w}}\right)^3}{\left(\ds{1^{w}2^{w}}-\da{1^{w}2^{w}}\right)\left(\ds{3^{w}1^{w}}+\da{3^{w}1^{w}}\right)}\prod_i\hat{w}_1^{I_i}\hat{u}^{J_i}_2\hat{u}^{K_i}_3\label{massivePT}
\end{align}
It is clear that the prefactor multiplying the spinors is the massive upgrade of the Parke-Taylor factor. The remaining factor accounts for the spin components with respect to a given quantisation axis. Likewise, the cross terms are of the form
\begin{align}
&\left(\frac{|u_2||u_3|}{|u_1|}\frac{|u_3||u_1|}{|u_2|}\hat{w}_1^{I_1}\hat{u}_1^{I_2}\hat{u}^{J_1}_2\hat{w}_2^{ J_2}\hat{u}^{K_1}_3\hat{u}^{K_2}_3\right)=\nonumber\\
&\qquad\qquad\frac{\left(\ds{2^{w}3^{w}}-\da{2^{w}3^{w}}\right)\left(\ds{3^{w}1^{w}}+\da{3^{w}1^{w}}\right)}{\ds{1^{w}2^{w}}-\da{1^{w}2^{w}}}\hat{w}_1^{I_1}\hat{u}_1^{I_2}\hat{u}^{J_1}_2\hat{w}_2^{J_2}\hat{u}^{K_1}_3\hat{u}^{K_2}_3.\label{massiveScalarVector}
\end{align}
The prefactor here suggestively resembles the massless amplitude for photon/gluon emission by a scalar. 

In all expressions prior to (\ref{3particle2}), all occurrences of the $\omega_i$ parameters cancelled-out and could be set to zero without loss of generality (effectively setting $\hat{w}_i=\hat{u}_i^\dagger$). While yielding the pleasing expressions above, the cost of the frame choice that sets $C=0$ is that further complication in the general expression has been transferred into the $\hat{w}_i$, which cannot be identified as unit spinors determined by the $\hat{u}^\dagger_i$ alone. Alternative expressions with this interpretation could be extracted directly from (\ref{3particle2UnFixed}) at the expense of additional manifest complication.

In the limit that all legs become massless, (\ref{non0fact}) implies that the frame spinors all converge to a particular helicity, which corresponds to the configuration of massless special $3$-particle kinematics. Either $\hat{u}_{i+}\rightarrow 0$ for each $i$ and the right-handed massless spinors align or $\hat{u}_{i-}\rightarrow 0$ and the left-handed spinors align. The combinations of bilinears (\ref{non0fact}) appearing in the superamplitude behave as $\left(\ds{i^{w}j^{w}}\pm\da{i^{w}j^{w}}\right)\rightarrow \hat{u}^\dagger_{i+}\hat{u}^\dagger_{j+}\ds{ij}$ or $\left(\ds{i^{w}j^{w}}\pm\da{i^{w}j^{w}}\right)\rightarrow \pm\hat{u}^\dagger_{i-}\hat{u}^\dagger_{j-}\da{ij}$ for each $i,j$. The surviving factors of $u_{iI}$ then become ``square-roots'' of the massless bilinears e.g. $u_{3+}\rightarrow\sqrt{\frac{\ds{23}\ds{31}}{\ds{12}}}$ or $u_{3+}\rightarrow\sqrt{\frac{\da{23}\da{31}}{\da{12}}}$. 

Furthermore, in either massless complex kinematical configuration, $C\rightarrow 0$ and only the terms retained in the $C=0$ frame remain in the massless limit. The factors of $\hat{w}_i$ may be identified with as $\hat{u}_i^\dagger$ and consequently the massless limit may be read-off from the expressions (\ref{3particle2}) and (\ref{3massivevecComp}). For example, in the case where all left-handed spinors become proportional, the second factor in (\ref{3particle2}) converges to $\delta^{(4)}(Q)$, while the first becomes the remaining Grassmann quadratic (including the Parke-Taylor factor) in the $\overline{\text{MHV}}$ term in (\ref{3massless}). The diagonal term in the all vector component amplitude above (\ref{massivePT}) clearly converges to the Parke-Taylor three vector amplitude $A[g_1^\mp,g_2^\pm,g_3^\pm]$ for the relevant helicity and massless special kinematics choices, otherwise it converges to zero. Likewise, the cross-terms like (\ref{massiveScalarVector}) converge to amplitudes expected for a Goldstone boson emitting a gluon, as expected from the Higgs mechanism. The remaining factors of the unit frame spinors $\hat{u}_{i\pm}^{(\dagger)}$ ultimately cancel-out.

In practice, although carrying the redundant reference spinor, the form (\ref{3massive}) is relatively easy to use in practical calculations. We will choose to continue to use the spacetime spinor formulation of (\ref{3massive}) in the remainder of this paper. Following a similar argument to that presented above for (\ref{3massive}), a simple representation for the $2$-massive-leg superamplitude may instead be derived by finding the SUSY invariant $\delta^{(4)}(Q^{\dagger a})\delta^{(4)}(Q_{a+2})$ and dropping one of the repeated factors of the degenerate supercharges. The overall coefficient is then fixed by the little group scaling of the legs. In this case, the special kinematics implies that $\la{3}p_1\rs{Q_{a+2}}=-m\da{3Q^{\dagger a}}$. The $3$-leg superamplitude is then determined to be 
\begin{align} \label{eqn:coulombbranch3pt}
\mathcal{A}_3[\mathcal{W}_1,\overline{\mathcal{W}}_2,G_3]&=\frac{-x}{m^3\da{q3}^2}\delta^{(4)}\left(Q^{\dagger a}\right)\delta^{(2)}\left(\la{q}p_1\rs{Q_{a+2}}\right)\nonumber\\
&=\frac{-x}{m^3\da{q3}^4}\delta^{(2)}\left(\da{3Q^{\dagger a}}\right)\delta^{(2)}\left(\da{qQ^{\dagger a}}\right)\delta^{(2)}\left(\la{q}p_1\rs{Q_{a+2}}\right)\nonumber\\
&=\frac{-x}{m\da{q3}^2}\delta^{(4)}\left(Q_{a+2}\right)\delta^{(2)}\left(\da{qQ^{\dagger a}}\right).
\end{align}
The reference spinor $\ra{q}$ must satisfy $\da{q3}\neq 0$. Because this superamplitude must be invariant under the little group scaling of its legs, the helicity-carrying factor $x$ has been re-introduced. The presence of $x$ is expected because of its appearance in the component amplitudes like $A_3[W^{I_1I_2},\overline W^{J_1J_2},g^+]$ and it emerges in taking the massless limit $m_3\rightarrow 0$ of $u_{3,K}\rs{3^K}/(\da{q3^K}u_{3,K})$ in (\ref{3massive}), as explained previously above.

Explicitly expanding the delta functions gives
\begin{align}
\mathcal{A}_3[\mathcal{W}_1,\overline{\mathcal{W}}_2,G_3]=\frac{-1}{m} x\prod_a\Big(-\ds{32^I}\eta_{1M}^a\eta_{1}^{Ma}\eta_{2I}^a+\ds{31^I}\eta_{1I}^a\eta_{2M}^{a}\eta_{2}^{aM}+\ds{1^I2^J}\eta_{1I}^a\eta_{2J}^a\eta_{3}^a\nonumber\\
-x^{-1}\da{1^I2^J}\eta_{1I}^a\eta_{2J}^a\tilde{\eta}_{3}^{\dagger a}+\frac{1}{2}m\eta_{1M}^a\eta_{1}^{Ma}\eta_3^a+\frac{1}{2}m\eta_{2M}^a\eta_{2}^{Ma}\eta_3^a+\frac{1}{2}\frac{m}{x}\eta_{1M}^a\eta_{1}^{Ma}\tilde{\eta}_3^{\dagger a}\nonumber\\
+\frac{1}{2}\frac{m}{x}\eta_{2M}^a\eta_{2}^{Ma}\tilde{\eta}_3^{\dagger a}+\ds{1^I3}\eta_{1I}^a\eta_3^a\tilde{\eta}_3^{\dagger a}-\ds{2^I3}\eta_{2I}^a\eta_3^a\tilde{\eta}_3^{\dagger a}\Big),
\end{align}
which allows the components to be efficiently read off. Notably, the reference spinor introduced in the delta functions has completely disappeared and does not affect the components. 

The form of the two-equal-mass superamplitude makes clear that the interactions of BPS states with massless gauge bosons are monomials in $x$. In the above case, this has the physical interpretation of the BPS states having gyromagnetic ratio $g=2$ exactly (and likewise no anomalous electric quadrupole moment, as seen in the $\mathcal{N}=1$ case in \cite{HKT:2018a}). The different Lorentz structures of the couplings are fully protected by supersymmetry. Explicitly, we may extract the collection of such component amplitudes as 
\begin{align}
\mathcal{A}[\mathcal{W},\overline{\mathcal{W}},g^-] &= \frac{x}{m} \prod\limits_{a} \left( [1^I 2^J] \eta_{1I}^a \eta_{2J}^a + \frac{1}{2} m \eta_{1M}^a \eta_1^{Ma} + \frac{1}{2} m \eta_{2M}^a \eta_2^{Ma}\right) \\
\mathcal{A}[\mathcal{W},\overline{\mathcal{W}},g^+] &= \frac{1}{mx} \prod\limits_{a} \left( \langle1^I 2^J\rangle \eta_{1I}^a \eta_{2J}^a - \frac{1}{2} m \eta_{1M}^a \eta_1^{Ma} - \frac{1}{2} m \eta_{2M}^a \eta_2^{Ma}\right).
\end{align}

\subsection{Four Particle Superamplitudes}\label{sec:4leg}

Using massive super-BCFW, we next present a derivation of the general $4$-leg superamplitude for legs of arbitrary mass. In this case, the Grassmann dependence is entirely determined by the factor $\delta^{(4)}(Q^{\dagger,a})\delta^{(4)}(Q_{a+2})$. Thus, only the coefficient of the delta function need be calculated and this is fixed by any single component amplitude. While the expected form of the superamplitude is obvious and follows from supersymmetry, factorisation and the (trivial) spin of the external superfields, the following derivation will illustrate how these emerge from combining the on-shell $3$-leg amplitudes (\ref{3massive}). It will also provide a simple demonstration of the mechanics and use of massive super-BCFW.

We will calculate $\mathcal{A}_{4}[\mathcal{W}_1,\overline{\mathcal{W}}_2,\mathcal{W}_3,\overline{\mathcal{W}}_4]$. This colour-ordering implies that the masses obey the constraint $m_1+m_3=m_2+m_4$. The cases with different combinations of particles and anti-particles may be obtained by obvious modification. For any such superamplitude that respects colour neutrality of the broken gauge group, there will always be two consistent factorisation channels in which the on-shell, internal particle has mass given by the sum of the masses of the other legs on each subamplitude (weighted by the sign of their central charges).

As noted above, $4d$ massive super-BCFW may be obtained by dimensionally reducing that of massless $6d$ $\mathcal{N}=(1,1)$ SYM.  An analogous calculation of the $4$-leg superamplitude in $6d$ was performed in \cite{Dennen:2009vk}, supersymmetrising the computation in pure YM in \cite{Cheung:2009dc}. In $4d$, the special case of two massless legs have been previously calculated by \cite{Craig:2011ws} and \cite{Cachazo:2018}, for the simple case of an $U(N+M)\rightarrow U(N)\times U(M)$ breaking pattern (where there are two possible structures with consistent colour-ordering). The former used non-supersymmetric BCFW recursion applied to the component amplitude $A_4[W,\overline{W},g^+,g^+]$ to determine the kinematical coefficient of the delta functions, while \cite{Cachazo:2018} used the general CHY-like formula. These are a special case of our result.

First define generalized Mandelstam variables $s_{ij} = -(p_i + p_j)^2 - (m_i \pm m_j)^2$, where the masses are added if the lines have the same sign central charge and subtracted if opposite. For a general amplitude with any number of legs, these satisfy the useful identities $\sum_j s_{ij}=0$ and $\sum_{j\neq k} s_{ij}=\sum_{j\neq i}s_{kj}$, by conservation of momentum and the mass constraint. Other relations may be similarly derived.

Applying the super-shift to legs $1$ and $2$, the superamplitude is determined from a single factorisation channel:
\begin{figure}[h]
\begin{center}
$\begin{gathered}
\begin{fmffile}{BCFW4}
\begin{fmfgraph*}(100,50)
     \fmfleft{i2,i1}
     \fmfright{o2,o1}
     \fmfset{arrow_len}{2mm}
     \fmf{fermion,label=$\hat{1}$,label.side=right,tension=5}{v1,i1}
     \fmf{fermion,label=$\bar{4}$,label.side=right,tension=5}{i2,v1}
     \fmf{fermion,label=$\hat{\bar{2}}$,label.side=right,tension=5}{o1,v2}
     \fmf{fermion,label=$3$,label.side=right,tension=5}{v2,o2}
     \fmfv{decor.shape=circle,decor.filled=empty,decor.size=20,label=$L$,label.dist=0}{v1}
     \fmfv{decor.shape=circle,decor.filled=empty,decor.size=20,label.dist=0,label=$R$}{v2}
     \fmf{fermion,label=$\hat{P}$,label.side=left,tension=5}{v1,v2}
     \end{fmfgraph*}
\end{fmffile}
\end{gathered}$
\end{center}
\caption{The single BCFW diagram for four-point recursion.}
\label{4legdiagram}
\end{figure}
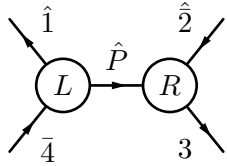

\begin{align}
\mathcal{A}_{4}[\mathcal{W}_1,\overline{\mathcal{W}}_2,\mathcal{W}_3,\overline{\mathcal{W}}_4]=\int d^4\eta_{\hat{P}}\hat{\mathcal{A}}_L[\overline{\mathcal{W}}_4,\widehat{\mathcal{W}}_1,\mathcal{W}_{\hat{P}}]\frac{-1}{s_{14}}\hat{\mathcal{A}}_R[\overline{\mathcal{W}}_{-\hat{P}},\widehat{\overline{\mathcal{W}}}_2,\mathcal{W}_3],
\end{align}
where $\hat{P}$ is the momentum of the internal line, taken as outgoing from the left subamplitude in Figure \ref{4legdiagram} and incoming into the right subamplitude. Hats denote shifted legs, to be evaluated on the residue determined by $s_{4\hat{1}}=0$, although it will be unnecessary in this example to determine either the residue or the shift vector. Assuming that $m_1<m_4$, then the internal on-shell particle has mass $m_{P}=m_4-m_1$ and is BPS in the left superamplitude. In the right amplitude, it is an incoming BPS state, which can be regarded by crossing symmetry as an outgoing anti-BPS state with momentum $-P$.

Analytically continuing spinors and Grassmann variables from negative to positive energies requires the rules
\begin{align}
\rs{-P^I}=i\rs{P^I}\qquad \ra{-P^I}=i\ra{P^I}\qquad \eta_{-P}^I=i\eta_{P}^I,
\end{align}
which are little group covariant (and consistent with \cite{Srednicki:2007qs}). These rules imply that the sign of the mass as it appears in the Weyl equation or the spin sums effectively reverses so that e.g. $p\rs{-p^I}=-m\ra{-p^I}$ for a leg of mass $m$ and momentum $p$. See Appendix A of \cite{HKT:2018a} for spinor conventions and identities. As a result, while the BPS condition for an analytically continued leg (\ref{BPS0}) is unchanged (noting that both the momentum and central charges reverse under crossing), a relative negative sign appears in the spinor-stripped counterparts (\ref{bpslimit}). As a result, under the conventions employed here, the left-handed multiplicative supercharges of the crossed legs pick up an extra negative sign relative to that of the other outgoing legs.

Likewise, the corresponding massless variables in the conventions employed here must also all acquire a factor of $i$ upon analytic continuation
\begin{align}\label{MasslessAnaCont}
\rs{-p}=i\rs{p}\qquad \ra{-p}=i\ra{p}\qquad \eta_{-p}=i\eta_{p} \qquad \eta_{-p}^\dagger=i\eta_{p}^\dagger.
\end{align}

The remaining calculation involves combining the delta functions and simplifying. It is through combining the delta functions that the extra pole is generated, effectively as a Jacobian factor arising from the mismatch between the aligned frame spinors $u_{\hat{P}}^I$ on the left and right on-shell amplitudes. This overlap was also the source of the additional pole in $6d$ YM \cite{Dennen:2009vk}. We give details of this in Appendix \ref{Appendix4leg}, but the result is that 
\begin{multline}
\hat{\mathcal{A}}_L[\overline{\mathcal{W}}_4,\widehat{\mathcal{W}}_1,\mathcal{W}_{\hat{P}}]\hat{\mathcal{A}}_R[\overline{\mathcal{W}}_{-\hat{P}},\widehat{\overline{\mathcal{W}}}_2,\mathcal{W}_3]=\frac{1}{\left(u_{\hat{P}M}^{(L)}u_{\hat{P}}^{(R)M}\right)^2}\frac{1}{(\la{q}p_4\hat{p}_1\ra{q})^2}\\
\times\delta^{(4)}\left(Q\right)\delta^{(4)}\left(Q^\dagger\right) \delta^{(2)}\left(\da{q\hat{Q}^\dagger_R}\right)\delta^{(2)}\left(\la{q}\hat{p}_1\rs{\hat{Q}_R}\right).
\end{multline}
Here the $L$ and $R$ subscripts index parameters originating in the factorised on-shell amplitudes, the hats indicate that they are shifted and the $U(2)$ $R$-indices have been omitted for brevity. The supercharges without subscripts represent those for the full $4$-leg superamplitude. The multiplicative factor arises from a succession of basis changes and invocations of constraints from the other delta functions. This critically provides the factor that will become the pole representing the other factorisation channel.

After factoring out the total supersymmetric delta function, the remaining Grassmann integral is simple to perform, giving
\begin{align}
\int d^4\eta_{\hat{P}}\delta^{(2)}\left(\da{q\hat{Q}^\dagger_R}\right)\delta^{(2)}\left(\la{q}\hat{p}_1\rs{\hat{Q}_R}\right)=\left(\la{q}\hat{p}_1\hat{P}\ra{q}\right)^2.
\end{align}
Combining all of the factors and using $\hat{P}=-\hat{p}_1-p_4$, the superamplitude reduces to
\begin{align}
\mathcal{A}_{4}[\mathcal{W}_1,\overline{\mathcal{W}}_2,\mathcal{W}_3,\overline{\mathcal{W}}_4]=\frac{\delta^{(4)}(Q)\delta^{(4)}(Q^\dagger)}{s_{41}}\frac{-1}{\left(u_{\hat{P}M}^{(L)}u_{\hat{P}}^{(R)M}\right)^2}.
\end{align}
All of the kinematical factors cancel out with the exception of the internal propagator for this factorisation channel and another factor given by the overlap of the frame spinors for the internal line. This form was also reached in the analogous $6d$ calculation \cite{Cheung:2009dc,Dennen:2009vk} and the demonstration that this is the pole of the other factorisation channel is similar. We repeat the argument from the $4d$ perspective in Appendix \ref{Appendix4leg}, the result of which is that 
\begin{align}
\left(u_{\hat{P}M}^{(L)}u_{\hat{P}}^{(R)M}\right)^2=-s_{12}.\label{MagicPole}
\end{align}
As explained in Subsection \ref{Sec:3leg}, the $u_i$ spinors select a preferred decomposition of massive momenta into a sum of two parallel null vectors. This new pole occurs when the frame spinors for the internal line in the BCFW diagram align. This equivalently means that the two sets of parallel null vectors that span the massive momenta on each side of the factorisation channel align. This is just the complexification of the alignment of the external massive momenta on opposite sides of the the factorisation in the BCFW diagram, which is exactly the condition required for the alternative factorisation channel.

The $4$-leg superamplitude is therefore
\begin{equation}
\mathcal{A}_{4}[\mathcal{W}_1,\overline{\mathcal{W}}_2,\mathcal{W}_3,\overline{\mathcal{W}}_4]=\frac{\delta^{(4)}(Q^{\dagger a})\delta^{(4)}(Q_{a+2})}{s_{12}s_{41}}.
\end{equation}
The residue and the momentum shift ultimately cancelled out in this calculation and did not have to be solved for. This amplitude closely resembles its counterparts in unbroken Yang-Mills, where both manifestly feature poles in both $s$ and $t$ factorisation channels in a single term. The superamplitude in which any of the legs is massless may be obtained as an obvious limiting case.

Just as in massless (super-)Yang-Mills, only one factorisation channel (or BCFW diagram) was sufficient to determine the $4$-leg amplitude from the elementary $3$-leg amplitudes. The pole providing the other factorisation channel originates from the kinematic ``singularity'' in the $3$-leg superamplitude (\ref{3massive}). The numerator (specified here by supersymmetry) determines the polarisation structure. It is a simple task to extract component amplitudes of massive states. As an example, the four massive vector boson amplitude may be found as
\begin{multline}\label{4massivecomp}
A_4[W^{I_1I_2}_1,\overline{W}^{J_1J_2}_2,W^{K_1K_2}_3,\overline{W}^{L_1L_2}_4]=\frac{1}{s_{12}s_{41}}\\
\prod_{i=1}^2\bigg(\ds{1^{I_i}2^{J_i}}\da{3^{K_i}4^{L_i}}+\da{1^{I_i}2^{J_i}}\ds{3^{K_i}4^{L_i}}+\ds{1^{I_i}3^{K_i}}\da{2^{J_i}4^{L_i}}\\+\da{1^{I_i}3^{K_i}}\ds{2^{J_i}4^{L_i}}+\ds{1^{I_i}4^{L_i}}\da{2^{J_i}3^{K_i}}+\da{1^{I_i}4^{L_i}}\ds{2^{J_i}3^{K_i}}\bigg)
\end{multline}
The massive little group indices are implicitly symmetrised over in the above expressions as usual. From the perspective of $6d$ Yang-Mills amplitudes dimensionally reduced to $4d$, each factor in the numerator is the reduction of the `$4$-bracket' of $6d$ spinors \cite{Cheung:2009dc}. It is clear that the expected helicity selection rules emerge in the massless limit (see Appendix A of \cite{HKT:2018a}), where the amplitudes without split helicities are mass suppressed, most severely when all helicities are the same.

\subsection{Five Particle Superamplitudes and Band Structure}\label{sec:5ptCoulomb}

\subsubsection{Bands}

Away from the origin of moduli space, the $R$-symmetry is broken to $USp(4)$ from $SU(4)$ and the sectors of distinct levels of helicity violation partially merge. This occurs because processes forbidden by helicity selection rules may now proceed at mass-suppressed rates. Instead, \cite{Craig:2011ws}, who work in a chiral superspace in which a $SU(2)\times SU(2)\leq USp(4)$ is manifest, are able to classify the residual supersymmetric invariant sectors by their Grassmann orders under each $SU(2)$ factor. Each of these sectors, in their formulation, is an inhomogeneous polynomial that spans several overlapping even Grassmann orders, which were described as `bands'. The polynomial of $(K+1)$th lowest degree was called the $\text{N}^K\text{MHV}$ sector, in analogy with the massless superamplitudes. Each invariant term in the superamplitude is then classified under this product structure as a $\text{N}^{k_1}\text{MHV}\times \text{N}^{k_2}\text{MHV}$ band.

In non-chiral superspace the superamplitudes are instead homogeneous of degree $2n$ in Grassmann variables and a distinct $U(2)$ $R$-subgroup is realized explicitly, which foretell a different organization of the bands here (we retain the term `band' for supersymmetrically closed sector, as well as the $\text{N}^{k_1}\text{MHV}\times \text{N}^{k_2}\text{MHV}$ notation). The simplest non-trivial example of a superamplitude with independent (albeit simple) bands is at five legs and our exploration here will provide insight into the band structure for general Coulomb branch superamplitudes. The three and four leg superamplitudes discussed above are special cases.

As discussed in \cite{Craig:2011ws}, the three leg massive superamplitude is non-trivial and actually contains three such independent terms. In the little group violating chiral superspace used by the authors, these appear as $\text{MHV}$ and $\overline{\text{MHV}}$ superamplitudes with a form almost identical to their massless counterparts, as well as a new $\text{MHV}\times\overline{\text{MHV}}$ term that vanishes in the massless limit. None of these are manifestly visible in our expression (\ref{3massive}), because they are represented by sectors of specific helicities, all of which are combined here into a massive little group invariant. That the three-leg superamplitude combines each helicity-violating band into a single, little group and supersymmetric invariant means that super-BCFW recursion cannot be automatically applied sector-by-sector as it is in the massless case. This weakening of the massless helicity selection rules may potentially complicate calculations if little group invariance is to be preserved.

We here illustrate the decomposition into bands of the 3-particle superamplitude, choosing the special case (\ref{eqn:coulombbranch3pt}) for simplicity. To reveal the separate supersymmetric invariant sectors, we explicitly strip off a massive spinor from one of the supercharges. We define 
\begin{equation}
\zeta^a_{1I} \equiv \frac{1}{m} \da{1_I Q^{\dagger a}}
\end{equation}
such that we may write the degenerate component of the delta functions as
\begin{equation}
\delta^{(2)}\left(\da{3Q^{\dagger a}}\right) = \delta^{(2)}\left(\da{31^I}\zeta^a_{1I}\right).
\end{equation}
The distinct bands closed under supersymmetry now correspond to the components of this sum in the helicity basis, so we may exhibit the band structure as
\begin{align}
\mathcal{A}_3[\mathcal{W}_1,\overline{\mathcal{W}}_2,G_3]&=\frac{-x}{m^3\da{q3}^4}\delta^{(2)}\left(\da{qQ^{\dagger a}}\right)\delta^{(2)}\left(\la{q}p_1\rs{Q_{a+2}}\right)\\
& \times \half \epsilon_{ab} \left[\da{31^+}^2\zeta^a_{1+}\zeta^b_{1+} +2 \da{31^+}\da{31^-}\zeta^a_{1+}\zeta^b_{1-} + \da{31^-}^2\zeta^a_{1-}\zeta^b_{1-}\right]\nonumber,
\end{align}
where the first term corresponds to the $\overline{\text{MHV}}$ band, the last to the MHV band, and the middle to the $\text{MHV}\times\overline{\text{MHV}}$ band, which vanishes in the massless limit. In other little group frames these bands will be scrambled, though still exist as separate supersymmetric invariants. This decomposition into bands makes it clear the way in which the separate sectors of helicity violation are combined in the massive case.

The four leg superamplitude has only one distinct supersymmetric structure. Just as for the massless case, there is only the $\text{MHV}$ sector, which is identical to its parity conjugate $\overline{\text{MHV}}$ sector. 

Beyond $4$ legs, the bands may be identified by solving the SWIs directly. At five legs, supersymmetry implies that $\mathcal{A}_5=\delta^{(4)}(Q^{\dagger,a})\delta^{(4)}(Q_{a+2}) F$, where $F$ is some function quadratic in Grassmann variables. Proceeding as in the general strategy laid out in \cite{HKT:2018a}, the appearance in $F$ of Grassmann variables for two of the lines may be eliminated here using the constraints imposed by the supersymmetric delta functions. Then supersymmetry requires that $Q_aF=0$ and $Q^{\dagger a+2}F=0$. These Grassmann PDEs may be solved by finding `Grassmann characteristics' - combinations of Grassmann variables upon which $F$ cannot depend. Then $F$ is a function of the other independent Grassmann variables that `label' the characteristics (this resembles the method used in \cite{Elvang:2009wd} to solve the SWIs). In this manner one may construct linear combinations of Grassmann variables, which we term `triads', that are annihilated by $Q_a$ and $Q^{\dagger a+2}$ and which include the $\eta$s of only three of the legs. 

Choosing a BPS line $i$ and an anti-BPS line $j$, we define Grassmann triads `anchored' at massless legs $k$ and massive legs $\ell$ as
\begin{align}\label{eqn:triads}
\xi^{a}_{k,ij} &\equiv \eta^a_k + \left(m_j \eta_{iI}^a \la{i^I} + m_i \eta_{jJ}^a \la{j^J}\right) \pi_{ij} \rs{k}/\pi_{ij}^2 \\
\tilde{\xi}^{\dagger a}_{k,ij} &\equiv \tilde{\eta}^{\dagger a}_k + \left(m_i \eta_{jJ}^a \ls{j^J} - m_j \eta_{iI}^a \ls{i^I} \right) \pi_{ij} \ra{k}/\pi_{ij}^2 \\
\xi^{a}_{\ell,ij L} &\equiv \eta^{a}_{\ell L} + \left(m_j \eta_{iI}^a \la{i^I} + m_i \eta_{jJ}^a \la{j^J}\right) \pi_{ij} \rs{\ell_L}/\pi_{ij}^2 \pm \left(m_i \eta_{jJ}^a \ls{j^J} - m_j \eta_{iI}^a \ls{i^I} \right) \pi_{ij} \ra{\ell_L}/\pi_{ij}^2 
\end{align}
where the upper sign in the last line is for BPS states and the lower sign for anti-BPS states, and we have defined $\pi_{ij} \equiv m_i p_j + m_j p_i$ for ease of reference. Since $i,j$ differ in the signs of their central charges, $\pi_{ij}^2 = m_i m_j s_{ij}$, where $s_{ij}$ are the generalized Mandelstam variables. The triads have the massless limits
\begin{align}
\xi^{a}_{k,ij} &\rightarrow \eta_k^a + \frac{\ds{jk}}{\ds{ij}} \eta_i^a + \frac{\ds{ki}}{\ds{ij}} \eta_j^a \equiv \frac{m^a_{ijk}}{\ds{ij}}\\
\tilde{\xi}^{\dagger a}_{k,ij} &\rightarrow \tilde{\eta}^{\dagger a}_k + \frac{\da{jk}}{\da{ij}} \tilde{\eta}^{\dagger a}_i + \frac{\da{ki}}{\da{ij}} \tilde{\eta}^{\dagger a}_j \equiv \frac{\tilde{m}^{\dagger a}_{ijk}}{\da{ij}} \\
\left(\xi^{a}_{\ell,ij+}, \xi^{a}_{\ell,ij-}\right) &\rightarrow \left(  \pm \frac{\tilde{m}^{\dagger a}_{ij\ell}}{\da{ij}},\frac{m^a_{ij\ell}}{\ds{ij}} \right),
\end{align}
where the massless $m^a_{ijk}=\ds{ij}\eta^a_k+\ds{jk}\eta^a_i+\ds{ki}\eta^a_j$ variables were recognized in \cite{Elvang:2009wd} as useful for solving the SWIs in the chiral superspace at the origin of moduli space. It is straightforward to take limits where only line $i$ or $j$ becomes massless. In the following we will use the same symbols for triads regardless of the masses of lines $i,j$, and rely on these limits to provide their definitions.

We may now write any superamplitude as a sum of a large-enough set of products of these triads with undetermined coefficients, and then project onto various component amplitudes to fix them. For a 5-leg superamplitude with up to four massive legs, we may characterize the band structure using the triads of a single massless leg as   
\begin{multline} \label{eqn:masslesschar}
\mathcal{A}_5\left[G_1,V_2,V_3,V_4,V_5\right] = \frac{\delta^{(4)}(Q^{\dagger,a})\delta^{(4)}(Q_{a+2})}{2 s_{45}^2}\epsilon_{ab} \times \bigg[ A_5[g^-,V_-,V_-,V_+,V_+] \xi^a_{1,23} \xi_{1,23}^b + \\ 2 A_5[S_{42}+S_{31},V_-,V_-,V_+,V_+] \xi^a_{1,23} \tilde{\xi}^{\dagger b}_{1,23} + A_5[g^+,V_-,V_-,V_+,V_+] \tilde{\xi}^{\dagger a}_{1,23} \tilde{\xi}^{\dagger b}_{1,23} \bigg].
\end{multline}
Here $V$ is either a massless $G$ or a massive $\mathcal{W}$ or $\overline{\mathcal{W}}$, while $V_\pm$ is the highest- or lowest-weight state in the multiplet, which are respectively $S_{12}, S_{34}$ and $\tilde{\phi},\phi$ for the massless and massive vectors. We note that the denominator merely cancels out the kinematic factors in the delta function and is not a pole, as the kinematic poles are contained within the component amplitudes which are here left undetermined. 

It is clear in this form that each of the terms is closed under supersymmetry. In the language of \cite{Craig:2011ws}, the first term in (\ref{eqn:masslesschar}) is the $\text{MHV}\times \text{MHV}$ band, the third is its parity conjugate and the second is the $\text{MHV}\times\overline{\text{MHV}}$ band (and its conjugate). Notably, this characterization of the bands respects little group covariance, but is determined by the massless multiplet's helicity states.

However, we may alternatively characterize the band structure using the triads of a single massive leg, which identifies the bands with the polarizations of the massive $W$. The 5-leg superamplitude with at least one massive leg may be written as 
\begin{equation} \label{eqn:massivechar}
\mathcal{A}_5\left[\mathcal{W}_1,V_2,V_3,V_4,V_5\right] = \frac{\delta^{(4)}(Q^{\dagger,a})\delta^{(4)}(Q_{a+2})}{2 s_{45}^2} A_5[W^{(IJ)},V_-, V_-,V_+,V_+] \epsilon_{ab}\,\xi^{a}_{1,23I}\xi_{1,23J}^b.
\end{equation}
The comparison of (\ref{eqn:masslesschar}) and (\ref{eqn:massivechar}) thus reflects clearly how the introduction of masses combines amplitudes of different helicity components and how this in turn combines the different bands of the superamplitude. 

As is evident in these formulae, the 5-leg superamplitudes have the special property that the bands are each fixed by a single component amplitude, so they may be fully determined once these are known. For the case of two massive legs, \cite{Craig:2011ws} used BCFW recursion to derive the partial amplitudes for a massive vector boson, its antiparticle and any number of massless gluons, which, after conversion to the little group covariant notation, may be written as 
\begin{multline}
A_n[W^{I_1I_2}_1,\overline{W}^{J_1J_2}_2,g^+_3,\ldots g^+_n]\\=\frac{-\da{1^{I_1}2^{J_1}}\da{1^{I_2}2^{J_2}}\ls{3}\prod_{i=4}^{n-1}(m^2-(p_i+\dots+p_n+p_1)(p_2+\dots+p_i))\rs{5}}{\da{34}\da{45}\dots\da{n-1 n}\prod_{i=4}^n((p_2+\dots+p_{i-1})^2+m^2)},\label{Allgplus}
\end{multline}
where $n-2$ is the number of gluon legs. Likewise, partial amplitudes with any number of massless scalars $S_{24}$ were derived as
\begin{align}
A_n[W^{I_1I_2},\overline{W}^{J_1J_2},S_{24},\ldots S_{24}]=\frac{m^{n-4}\ds{1^{I_1}2^{J_1}}\da{1^{I_2}2^{J_2}}}{\prod_{i=4}^n((p_2+\dots+p_{i-1})^2+m^2)},\label{Allscalar}
\end{align}
where $n-2$ is the number of scalar legs. 

For $5$-legs, these component amplitudes may be combined into the superamplitude
\begin{multline}\label{eqn:5ptexample}
\mathcal{A}_{5}[\mathcal{W}_1,\overline{\mathcal{W}}_2,G_3,G_4,G_5]=\frac{\delta^{(4)}(Q^{\dagger,a})\delta^{(4)}(Q_{a+2})}{s_{51}s_{23}s_{45}}\epsilon_{ab}\times\\
\left(\frac{\la{3}p_2p_1-m^2\ra{5}}{2\ds{34}\da{45}}\xi^a_{3,12} \xi^b_{3,12}+m\xi^a_{3,12}\tilde{\xi}^{\dagger b}_{3,12} +\frac{\ls{3}p_2p_1-m^2\rs{5}}{2\da{34}\ds{45}}\tilde{\xi}^{\dagger a}_{3,12}\tilde{\xi}^{\dagger b}_{3,12}\right),
\end{multline}
where
\begin{align} \label{eqn:Grassmannvars}
\xi^a_{3,12}&=\frac{-1}{s_{12}}\left(\ls{3}p_1+p_2\ra{1^I}\eta^a_{1I}+\ls{3}p_1+p_2\ra{2^J}\eta^a_{2J}+s_{12}\eta^a_3\right)\nonumber\\
\tilde{\xi}^{\dagger a}_{3,12}&=\frac{-1}{s_{12}}\left(\la{3}p_1+p_2\rs{1^I}\eta^a_{1I}-\la{3}p_1+p_2\rs{2^J}\eta^a_{2J}+s_{12}\tilde{\eta}^{\dagger a}_3\right).
\end{align}
Note that the denominator of the superamplitude is somewhat different from (\ref{eqn:masslesschar}) as the component amplitudes that have been matched onto are different, but the band structure is still clearly visible in terms of orders in helicity violation. As anticipated, the $\xi^a_{3,12}\tilde{\xi}^{\dagger b}_{3,12}$ term, which represents the $\text{MHV}\times\overline{\text{MHV}}$ band, clearly vanishes in the massless limit, leaving the usual $\text{MHV}$ sector and its parity conjugate $\overline{\text{MHV}}$.

With more legs, each band can consist of multiple combinations of triads and they are also no longer fixed by single component amplitudes. The exceptions to this, most clearly illustrated if there are enough massless legs for the superamplitude to be described entirely with massless triads, are always the $\text{MHV}\times \text{MHV}$ band, which corresponds to the only term that is purely holomorphic in triads anchored at massless legs (and analogously for the $\overline{\text{MHV}}\times\overline{\text{MHV}}$ band), and the $\text{MHV}\times\overline{\text{MHV}}$ band, which involves a single term with an equal number of triads and conjugate triads, each of a different type. For example, the $6$-leg superamplitude $\mathcal{A}[\mathcal{W},\overline{\mathcal{W}},G,G,G,G]$ has bands described by $\xi^a_{3,45}$, $\xi^a_{4,56}$ and their conjugates. The $\text{MHV}\times \text{MHV}$ band is given by the single holomorphic term $\epsilon_{ab}(\xi^a_{3,45}\xi^b_{3,45})\epsilon_{cd}(\xi^c_{4,56}\xi^d_{4,56})$, the terms in the $\text{NMHV}\times \text{MHV}$ and $\text{MHV}\times \text{NMHV}$ bands are of the form $\sim\xi^3\tilde{\xi}$, while the terms in the $\text{NMHV}\times \text{NMHV}$ band are of the form $\sim\xi^2\tilde{\xi}^2$. However, when most of the legs are massive, there will not be a form in which all of the Grassmann triads are anchored to massless legs and the little group will combine the bands into components of an $SU(2)$ tensor, similar to that observed in (\ref{eqn:massivechar}).

In addition to having more available Grassmann structures, terms within each band are related by the massive $R$-symmetry generators (\ref{MassiveR}) that are not part of the $U(2)$ linearly represented on the on-shell superspace. A similar analysis to \cite{Elvang:2009wd} could be performed to determine the Grassmann structure for higher leg superamplitudes. We will instead return our attention toward super-BCFW recursion, which has the capacity to generate complete expressions instead.

\subsubsection{Five Particle Superamplitudes}

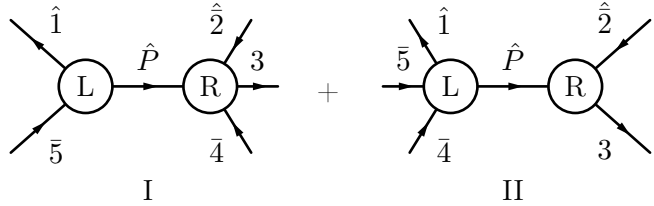
\begin{figure}[h]
\begin{center}
$\begin{gathered}
\begin{fmffile}{BCFW51}
\begin{fmfgraph*}(100,50)
	 \fmfstraight
     \fmfleft{i2,i1}
     \fmfcurved
     \fmfright{o3,o2,o1}
     \fmfset{arrow_len}{2mm}
     \fmf{fermion,label=$\hat{1}$,label.side=right,tension=5}{v1,i1}
     \fmf{fermion,label=$\bar{5}$,label.side=right,tension=5}{i2,v1}
     \fmf{fermion,label=$\hat{\bar{2}}$,label.side=right,tension=5}{o1,v2}
     \fmf{phantom,tension=1,label=I,label.dist=35}{v1,v2}
     \fmf{fermion,label=$\bar{4}$,label.side=left,tension=5}{o3,v2}
     \fmf{fermion,label=$3$,label.side=left,tension=5}{v2,o2}
\fmfv{decor.shape=circle,decor.filled=empty,decor.size=20,label=L,label.dist=0}{v1}
     \fmfv{decor.shape=circle,decor.filled=empty,decor.size=20,label.dist=0,label=R}{v2}
     \fmf{fermion,label=$\hat{P}$,label.side=left,tension=5}{v1,v2}
     \end{fmfgraph*}
\end{fmffile}
\end{gathered}$
\quad + \quad 
$\begin{gathered}
\begin{fmffile}{BCFW52}
\begin{fmfgraph*}(100,50)
	 \fmfcurved
     \fmfleft{i3,i2,i1}
     \fmfstraight
     \fmfright{o2,o1}
     \fmfset{arrow_len}{2mm}
     \fmf{fermion,label=$\hat{1}$,label.side=right,tension=5}{v1,i1}
     \fmf{fermion,label=$\bar{5}$,label.side=left,tension=5}{i2,v1}
     \fmf{fermion,label=$\bar{4}$,label.side=right,tension=5}{i3,v1}
     \fmf{fermion,label=$\hat{\bar{2}}$,label.side=right,tension=5}{o1,v2}
     \fmf{fermion,label=$3$,label.side=right,tension=5}{v2,o2}
     \fmf{phantom,tension=1,label=II,label.dist=35}{v1,v2}     \fmfv{decor.shape=circle,decor.filled=empty,decor.size=20,label=L,label.dist=0}{v1}
     \fmfv{decor.shape=circle,decor.filled=empty,decor.size=20,label.dist=0,label=R}{v2}

     \fmf{fermion,label=$\hat{P}$,label.side=left,tension=5}{v1,v2}
     \end{fmfgraph*}
\end{fmffile}
\end{gathered}$
\end{center}
\caption{The two BCFW diagrams for five-point recursion.}
\label{5legDiagram}
\end{figure}

Using the insight provided above into the helicity structure of the the $5$-leg superamplitude, we proceed to use massive super-BCFW to compute it in full generality. This gives a first non-trivial application of BCFW recursion to computing amplitudes in which every leg is massive. 

Much of the ensuing calculation resembles that performed in $6d$ in \cite{Dennen:2009vk} and \cite{Cheung:2009dc}. However, utilising the interpretation of the bands above, we are able to take short-cuts, despite the calculation presumably being attainable through dimensional reduction and not yet adapted with variables likely accommodating of dual conformal symmetry, as used in \cite{Dennen:2010dh} and \cite{Plefka:2014fta}.

We will choose to compute the superamplitude $\mathcal{A}[\mathcal{W}_1,\overline{\mathcal{W}}_2,\mathcal{W}_3,\overline{\mathcal{W}}_4,\overline{\mathcal{W}}_5]$ for $m_1<m_5$. Results for other choices of central charges and masses are obtained by trivial modification. Applying the massive super-BCFW shift to the first and second legs, the superamplitude recurses to the two factorisation channels depicted in Figure \ref{5legDiagram}. The resulting superamplitude is 
\begin{multline}
\mathcal{A}[\mathcal{W}_1,\overline{\mathcal{W}}_2,\mathcal{W}_3,\overline{\mathcal{W}}_4,\overline{\mathcal{W}}_5]=\int d^4\eta_{\hat{P}}\hat{\mathcal{A}}_L[\overline{\mathcal{W}}_5,\widehat{\mathcal{W}}_1,\mathcal{W}_{\hat{P}}]\frac{-1}{s_{15}}\hat{\mathcal{A}}_R[\overline{\mathcal{W}}_{-\hat{P}},\widehat{\overline{\mathcal{W}}}_2,\mathcal{W}_3,\overline{\mathcal{W}}_4]\Big |_{z_*^{(1)}}\\
+\int d^4\eta_{\hat{P}}\hat{\mathcal{A}}_L[\overline{\mathcal{W}}_4,\overline{\mathcal{W}}_5,\widehat{\mathcal{W}}_1,\mathcal{W}_{\hat{P}}]\frac{-1}{s_{23}}\hat{\mathcal{A}}_R[\overline{\mathcal{W}}_{-\hat{P}},\widehat{\overline{\mathcal{W}}}_2,\mathcal{W}_3]\Big |_{z_*^{(2)}}.
\end{multline}
Each term is to be evaluated upon a different pole, respectively determined to be at
\begin{align}
s_{\hat{1}5}=0&\Rightarrow z_*^{(1)}=\frac{s_{15}}{2r\cdot p_5}\nonumber\\
s_{\hat{2}3}=0&\Rightarrow z_*^{(2)}=\frac{-s_{23}}{2r\cdot p_3}.\label{residues}
\end{align}
In each term, combining the delta functions to produce the full $5$-leg supersymmetric delta function is easy, as, on the support of the $3$-leg superamplitude's delta function, the $4$-leg superamplitude's delta function is equivalent to the overall delta function for the full superamplitude. This leaves the $3$-leg delta function to be integrated in the state sum.

The calculation may be continued by substituting the shifted momenta and Grassmann variables into the two terms, adding them together and simplifying. Outside the total supercharge conserving delta functions, only $\eta_1$, $\eta_2$, $\eta_3$ and $\eta_5$ manifestly appear in the two terms above. However, we know that these must able to be arranged (after use of the constraints imposed by the delta functions) into the triad structure discussed above, such as that succinctly presented in (\ref{eqn:massivechar}). Choosing to represent the superamplitude using triads $\xi^a_{3,12}$, we only need to identify the coefficient of the terms containing a factor of $\epsilon_{ab}\eta_{3 K_1}^a\eta_{3K_2}^b$ to bootstrap the entire superamplitude. This is because, in this case, the massive bands are each determined by a polarisation component of a single component amplitude, $A[\phi_1,\phi_2,W^{K_1K_2}_3,\tilde{\phi}_4,\tilde{\phi}_5]$. This allows us to henceforth discard all terms in the calculation that do not have a factor of $\epsilon_{ab}\eta_{3 K_1}^a\eta_{3K_2}^b$, but only after first using the delta function constraints to eliminate $\eta_4$ and $\eta_5$, the latter of which appears in the first term. 

Inverting the delta function constraints $Q=0$ and $Q^\dagger=0$ for the supercharges implies that, in expressing $\ra{5^M}\eta_{5M}$ as a linear combination of $\eta_1$, $\eta_2$ and $\eta_3$, the latter term is
\begin{align}
\frac{m_4m_5}{s_{45}}\left(1-\frac{p_5p_4}{m_4m_5}\right)\left(\frac{p_4}{m_4}+\frac{p_3}{m_3}\right)\rs{3^K}\eta_{3K}=\frac{m_4m_5}{s_{45}}\ra{A^K}\eta_{3K},\label{5to3}
\end{align}
where we define the spinor $\ra{A^K}$ above to condense notation (we do not bother here to present the terms proportional to other Grassmann variables, as these do not contribute to the $\eta_3^2$ term in the superamplitude).

Resuming our calculation of the BCFW diagrams in Figure \ref{5legDiagram}, the first diagram contributes 
\begin{multline}
\int d^4\eta_{\hat{P}}\hat{\mathcal{A}}_L[\overline{\mathcal{W}}_5,\widehat{\mathcal{W}}_1,\mathcal{W}_{\hat{P}}]\frac{-1}{s_{15}}\hat{\mathcal{A}}_R[\overline{\mathcal{W}}_{-\hat{P}},\widehat{\overline{\mathcal{W}}}_2,\mathcal{W}_3,\overline{\mathcal{W}}_4]\Big |_{z_*^{(1)}}\\
=\frac{\delta^{(4)}(Q)\delta^{(4)}(Q^\dagger)}{s_{15}s_{\hat{2}3}s_{34}}\frac{1}{\la{q}p_5\hat{p}_1\ra{q}}\prod_a\left(m_1\da{q5^M}-\la{q}\hat{p}_1\rs{5^M}\right)\eta_{5M}^a\Big |_{z_*^{(1)}},\label{Term1}
\end{multline}
where we have performed the Grassmann intergral and retained only the $\eta_5^2$ terms. The exchange symmetry of the little group indices of the factors contracted against the $\eta_5$ variables implies that, by fermion statistics, only the component of the product of $\eta_5$ variables that is antisymmetric in $R$-indices provides a non-zero contribution, so $\eta_{5M_1}^1\eta_{5M_2}^2\sim\frac{-1}{2}\epsilon_{ab}\eta_{5M_1}^a\eta_{5M_2}^b$. Then applying the Schouten identity, (\ref{Term1}) can be simplified to
\begin{align}
\frac{\delta^{(4)}(Q)\delta^{(4)}(Q^\dagger)}{s_{15}s_{\hat{2}3}s_{34}}\frac{-1}{m_5^2}\la{5^{M_1}}\hat{p}_1p_5\ra{5^{M_2}}\left(\frac{-1}{2}\epsilon_{ab}\eta_{5M_1}^a\eta_{5M_2}^b\right)
\end{align}
(leaving implicit evaluation on the first residue). Substituting in (\ref{5to3}) gives the $\eta_3^2$ contribution from the first BCFW diagram
\begin{align}
\frac{\delta^{(4)}(Q)\delta^{(4)}(Q^\dagger)}{s_{15}s_{\hat{2}3}s_{34}s_{45}^2}(-m_4^2)\la{A^{K_1}}\hat{p}_1p_5\ra{A^{K_2}}\left(\frac{-1}{2}\epsilon_{ab}\eta_{3K_1}^a\eta_{3K_2}^b\right).
\end{align}
The spinor bilinear in the term above may be simplified to 
\begin{multline}
\la{A^{K_1}}\hat{p}_1p_5\ra{A^{K_2}}=\frac{1}{m_3m_4^2}\bigg(\la{3^{K_1}}p_4\rs{3^{K_2}}(s_{45}s_{\hat{1}3} -s_{\hat{1}4}s_{35}) \\ -\la{3^{K_1}}\hat{p}_1\rs{3^{K_2}}s_{45}s_{34}+\la{3^{K_1}}p_5\rs{3^{K_2}}s_{\hat{1}4}s_{34}\bigg).
\end{multline}

The second BCFW diagram in Figure \ref{5legDiagram} may be evaluated almost identically to the first. In this case no factors of $\eta_4$ or $\eta_5$ appear, so only the coefficient of the $\eta_3^2$ term needs to be retained. This contributes
\begin{align}
\frac{\delta^{(4)}(Q)\delta^{(4)}(Q^\dagger)}{s_{45}s_{23}s_{\hat{1}5}}\frac{-1}{m_3}\la{3^{K_1}}\hat{p}_2\rs{3^{K_2}}\left(\frac{-1}{2}\epsilon_{ab}\eta_{3K_1}^a\eta_{3K_2}^b\right)\Big |_{z_*^{(2)}}.
\end{align}

The next step is to add the two BCFW terms together and combine them into a simplified expression. Explicitly evaluated on the residues (\ref{residues}), the shifted Mandelstam invariants appearing in each term may be expressed as
\begin{align}
s_{\hat{2}3}\Big |_{z_*^{(1)}}=\frac{1}{r\cdot p_5}\left(s_{23}(r\cdot p_5)+s_{15}(r\cdot p_3)\right), \qquad s_{\hat{1}5}\Big |_{z_*^{(2)}}=\frac{1}{r\cdot p_3}\left(s_{23}(r\cdot p_5)+s_{15}(r\cdot p_3)\right).
\end{align}
Following \cite{Dennen:2009vk} by calling $\phi=s_{23}r\cdot p_5+s_{15}r\cdot p_3$, the two BCFW terms can be combined to give 
\begin{align}
-\frac{\delta^{(4)}(Q)\delta^{(4)}(Q^\dagger)}{m_3s_{23}s_{34}s_{45}s_{51}\phi}\Big(&\frac{s_{23}}{s_{45}}(r\cdot p_5)\Big(\la{3^{K_1}}p_4\rs{3^{K_2}}(s_{45}s_{\hat{1}3}-s_{\hat{1}4}s_{35})-\la{3^{K_1}}\hat{p}_1\rs{3^{K_2}}s_{45}s_{34}\nonumber\\
&+\la{3^{K_1}}p_5\rs{3^{K_2}}s_{\hat{1}4}s_{34}\Big)\Big|_{z_*^{(1)}}+s_{34}s_{15}(r\cdot p_3)\la{3^{K_1}}\hat{p}_2\rs{3^{K_2}}\Big|_{z_*^{(2)}}\Big).\label{PartialCombine}
\end{align}
The terms proportional to $\la{3^{K_1}}r\rs{3^{K_2}}$ arising from the shifted momenta cancel after substituting the residues. In order to progress further, the ambiguity from momentum conservation and the mass selection rule can be fixed to help combine terms. Choosing to do this by eliminating $p_4$ and $m_4$ from the expression, the special identities between Mandelstam invariants introduced just prior to the computation of the $4$-leg superamplitude in Section \ref{sec:4leg} can be used for simplification. In doing so, all remaining dependence on the shift vector $r$ in the numerator of (\ref{PartialCombine}) factorises into a factor of $\phi$ and thus cancels against that in the denominator. This leaves an expression for the $\eta_3^2$ term in the $5$-leg superamplitude that is independent of the shift vector and is little group covariant. The full superamplitude may then be obtained by the replacement $\eta_3\mapsto\xi_{3,12}$. The result is 
\begin{multline}\label{eqn:5ptsuperamp}
\mathcal{A}[\mathcal{W}_1,\overline{\mathcal{W}}_2,\mathcal{W}_3,\overline{\mathcal{W}}_4,\overline{\mathcal{W}}_5]=-\frac{\delta^{(4)}(Q)\delta^{(4)}(Q^\dagger)}{2m_3s_{51}s_{23}s_{34}s_{45}^2}\Big(s_{23}(s_{15}+s_{25})\la{3^{K_1}}p_1\rs{3^{K_2}}\\
-s_{23}s_{12}\la{3^{K_1}}p_5\rs{3^{K_2}}+(s_{35}s_{12}-s_{13}(s_{25}+s_{15}))\la{3^{K_1}}p_2\rs{3^{K_2}}\Big)\epsilon_{ab}\xi^a_{3,12 K_1}\xi^b_{3,12 K_2},
\end{multline}
which agrees precisely with (\ref{eqn:5ptexample}) in the appropriate limit. 

Remarkably, at no point in the calculation is the identity of the shift vector $r$ actually needed - it cancels out in the end. However, just as for the cancellation of spurious poles observed in massless recursion at higher legs, this only occurs after contributions from distinct factorisation channels (BCFW diagrams) are added together.  This means that, despite manifestly breaking little group covariance, recursion nevertheless delivers a little group invariant expression. While invariance is not manifest term-by-term, it is broken in a controlled way. The shift vector seems only to be the needle threading the factorisation channels into a complete superamplitude. This clearly invites a search for an alternative picture of how the factorisation channels are being combined. Especially important to be investigated is the significance of the little group breaking in the BCFW representation of the superamplitude for dual (super)conformal invariance.

Although the $5$-leg superamplitude does have non-trivial distinct bands, it is nevertheless an especially simple example in which each band is determined by a single supersymmetry invariant, and hence component amplitude, so that only the $U(2)$ $R$-symmetry subgroup provides non-trivial, independent constraints. This may be anticipated from its massless counterpart, which consists only of $\text{MHV}$ and a distinct, yet parity-conjugate, $\overline{\text{MHV}}$ sector. These are especially simple to derive using massless super-BCFW recursion. In the massive case considered here, the three different bands, most clearly visible in (\ref{eqn:5ptexample}) when some of the legs are massless, may be directly attributed to those of the three-leg superamplitude that are fused in super-BCFW recursion along the factorisation channel. We leave to be explored exactly how a massive manifestation of dual conformal invariance, which, for massless superamplitudes, is provided through super-BCFW, may interplay with both the little group and the band structure. Many simplifying features at five legs will not be present at six legs, which will provide a more acute test of the symmetries, their constraining power and the usefulness and meaning of recursion.

It was proposed by \cite{Huang:2011um} that the $6d$ SYM superamplitudes (or equivalently, the $4d$ massive superamplitudes) could be entirely determined (or ``uplifted'') by their restrictions to $4d$ massless states. This made use of the expressions using dual variables, in which both dual conformal and permutation symmetries can be made manifest with relatively simple expressions for the superamplitudes. The uplift was demonstrated up to $5$ legs, where the compact structure made it obvious by eye, once compact $4d$ and $6d$ building blocks manifesting the dual symmetries were identified. However, complications were encountered in \cite{Plefka:2014fta} at six legs, where the form of the $4d$ superamplitude produced by BCFW recursion in non-chiral superspace was not automatically amenable to the uplift. Again, the difference arises because of the new, independent bands and their additional structures. 

Because we are not fully manifesting the symmetries, in particular parity (through our use of chiral spinors) and the $6d$ Lorentz invariance, the uplift from (\ref{eqn:5ptexample}) (or its fully massless limit) to (\ref{eqn:5ptsuperamp}) is not obvious, although there is a clear resemblance in the structures, especially in the way that the bands are combined (the converse operation, the massless limit, is easily seen and verified and relates the terms in the two expressions). It is suggested in \cite{Plefka:2014fta} that the MHV sector by itself in the massless theory would be sufficient to determine the entire massive superamplitude, if the uplift were correct. This is plausible for the five leg case here, where the MHV sector corresponds to a single little group combination of the triads in (\ref{eqn:5ptsuperamp}). However, the continuation at six legs is the real test. The extent to which the embedded massless theory controls the structure of the massive theory remains an open question.

\section{Conclusion}

The spinor helicity formalism has provided a set of variables with respect to which a broad set of phenomena can be formulated and uncovered on-shell purely through recourse to fundamental principles of quantum mechanics and relativity, without introducing quantum fields and path integrals and their associated unphysical redundancies. The little group has provide an organisation of these variables enabling them to be adapted to insightfully describe the kinematics of massive particles. Treating both massive and massless states on the same footing, it may then be determined precisely to what extent features of quantum field theories are emergent from assumptions about infrared properties. Supersymmetry provides an idealisation that is already known to enhance many of the on-shell properties of unbroken Yang-Mills amplitudes. 

The power of on-shell methods for massive theories may be strongest on the Coulomb branch of $\mathcal{N}=4$ SYM, the maximally supersymmetric theory of massive particles, just as they are at the origin of moduli space for massless states. As a first step toward fully determining the on-shell properties of the theory, we have determined the elementary three-leg superamplitudes. These superamplitudes surprisingly have kinematic factors in their denominators akin to those of massless (super-)Yang-Mills, despite this not being a feature of their component amplitudes. Using super-BCFW recursion for amplitudes of massive particles, we have shown how, by combining on-shell $3$-particle superamplitudes across a factorisation channel, a new pole emerges that completes the $4$-leg massive superamplitude. This pole arises from combining supersymmetry invariants across the factorisation channel, an operation that simultaneously ensures that the arbitrary reference spinors in the $3$-particle superamplitudes are cancelled. This property is not a feature of the non-supersymmetric Higgsed Yang-Mills counterpart. We have then provided the first non-trivial use of BCFW recursion to compute a scattering amplitude entirely involving massive particles, doing so to determine the general $5$-leg superamplitude on the Coulomb branch.

The next objective is to compute higher leg superamplitudes. We have shown here that massive super-BCFW recursion offers an avenue for doing this. However, guidance is still necessary for interpreting the expressions that it leaves. Just as for the massless superamplitudes, such a beacon may be provided by dual conformal symmetry. Super-BCFW for massless amplitudes was crucial in deriving a representation in which the dual superconformal symmetry could be deduced (as a sum over $R$-invariants that, in momentum twistor space, makes dual conformal symmetry manifest). However, its full consequences for the massive amplitudes and relationship with the little group has yet to be fully elucidated. Also, while we have demonstrated that super-BCFW is indeed valid, we expect that, just as at the origin of moduli space, this is more directly a consequence (or maybe expression of) dual conformal symmetry or a deeper structure. The hypothesized Grassmannian formulation of the $\mathcal{N}=4$ SYM amplitudes on the Coulomb branch - the `symplectic Grassmannian' \cite{ArkaniHamed:2012nw}, may make this more explicit. The $6d$ point of view may also provide the framework within which these structures can be seen \cite{Cachazo:2018}, \cite{Geyer:2018xgb}.

Having established the on-shell properties of this idealised theory, the extent to which they descend to theories with less supersymmetry remains to be explored. In the massless theory, the constructibility of the superamplitudes descend to those of pQCD. We have given a brief discussion of how certain tree superamplitudes may be projected down to theories of less supersymmetry in Appendix \ref{sec:projecting}. Further progress would require a strategy for projecting out effectively closed subsectors or finding an adaptation of massive (super)-BCFW recursion to these theories. 

\acknowledgments 

We thank Tim Cohen, Nathaniel Craig and Henriette Elvang for comments on a draft of this work, Nathaniel Craig for discussions and support during the completion of this work, and Nima Arkani-Hamed and Yu-tin Huang for discussions on \cite{Arkani-Hamed:2017jhn}. AH and SK are grateful for the support of a Worster Fellowship. This work is supported in part by the US Department of Energy under the grant DE-SC0014129. 

\appendix

\section{Four Particle Superamplitude Details}\label{Appendix4leg}

To begin combining the delta functions in the $3$-leg superamplitudes, we express both left and right superamplitudes in the form in the first line of (\ref{3massive}). Clearly $\delta^{(4)}(\hat{Q}_L)\delta^{(4)}(\hat{Q}_R)=\delta^{(4)}(Q)\delta^{(4)}(\hat{Q}_R)$. By construction, $\hat{Q}_L+\hat{Q}_R=Q$ is unshifted (and similarly for the conjugate supercharges). Then representing the right delta function as the second line in (\ref{3massive}) and using $\delta^{(4)}(\hat{Q}_R)=\frac{1}{m_1^2\la{q}p_4\hat{p}_1\ra{q}}\delta^{(2)}\left(\la{q}\hat{p}_1\rs{\hat{Q}_R}\right)\delta^{(2)}\left(u_{4L}\la{4^L}\hat{p}_1\rs{\hat{Q}_R}\right)$ gives  $\delta^{(2)}\left(\la{q}\hat{p}_1\rs{\hat{Q}_L}\right)\delta^{(2)}\left(\la{q}\hat{p}_1\rs{\hat{Q}_R}\right)=\delta^{(2)}\left(\la{q}\hat{p}_1\rs{Q}\right)\delta^{(2)}\left(\la{q}\hat{p}_1\rs{\hat{Q}_R}\right)$. Note that the same reference spinor $\ra{q}$ may be used for both left and right factors. Such a spinor always exists that is parallel to neither $u_{4L}\la{4^L}$ nor $u_{3K}\la{3^K}$. This leaves the remaining delta functions $\delta^{(2)}\left(\da{q\hat{Q}^\dagger_R}\right)\delta^{(2)}\left(\la{q}\hat{p}_1\rs{\hat{Q}_R}\right)\delta^{(2)}\left(u_{4L}\la{4^L}\hat{p}_1\rs{\hat{Q}_R}\right)$ from which to extract the final factor required for the full $4$-leg delta function. 

On the combined support of the other delta functions, $\delta^{(2)}\left(u_{4L}\la{4^L}\hat{p}_1\rs{\hat{Q}_R}\right)$ \\ $= C^2\delta^{(2)}\left(u_{4L}\la{4^L}\hat{p}_1\rs{Q}\right)$, where the constant $C^2=\frac{m_1^2}{\left(u_{\hat{P}M}^{(L)}u_{\hat{P}}^{(R)M}\right)^2}\frac{\la{q}p_3\hat{p}_2\ra{q}}{\la{q}p_4\hat{p}_1\ra{q}}$. This follows from the relations
\begin{align}
u_{\hat{P}M}^{(R)}\la{\hat{P}^M}&=\alpha \,u_{\hat{P}M}^{(L)}\la{\hat{P}^M}+\beta\la{q}\\
u_{\hat{P}M}^{(R)}\ls{\hat{P}^M}&=\alpha \,u_{4L}\la{4^L}\frac{\hat{p}_1}{m_1}+\beta\la{q}\frac{\hat{P}}{m_P},
\end{align}
where
\begin{align}
\alpha=\frac{u_{3K}\da{3^Kq}}{u_{4L}\da{4^Lq}}\qquad\beta=\frac{u_{3K}\da{3^K4^L}u_{4L}}{\da{q4^L}u_{4L}}.
\end{align}
The scalar coefficients of the spinors in the first line above have been obtained by use of (\ref{M3PSK}). The second line may be obtained by (\ref{M3PSK}) and the Weyl equations. On the support of $\delta^{(2)}\left(\da{q\hat{Q}^\dagger_R}\right)$ and $\delta^{(4)}(Q^\dagger)\propto\delta^{(4)}\left(\da{q\hat{Q}_R^\dagger}+\da{q\hat{Q}_L^\dagger}\right)$, then $\da{q\hat{Q}_R^\dagger}, \da{q\hat{Q}_L^\dagger}\sim 0$ and terms proportional to these in the other delta functions may be dropped. Thus
\begin{align}
\delta^{(4)}(Q^\dagger)&\propto\delta^{(2)}(u_{\hat{P}M}^{(R)}\da{\hat{P}^MQ^\dagger})\nonumber\\
&=\delta^{(2)}(u_{\hat{P}M}^{(R)}\da{\hat{P}^M\hat{Q}^\dagger_R}+\alpha u_{\hat{P}M}^{(L)}\da{\hat{P}^M\hat{Q}^\dagger_L})\nonumber\\
&=\delta^{(2)}(u_{\hat{P}M}^{(R)}\ds{\hat{P}^M\hat{Q}_R}+\alpha u_{\hat{P}M}^{(L)}\ds{\hat{P}^M\hat{Q}_L})\nonumber\\
&=\delta^{(2)}\left(\alpha u_{4L}\la{4^L}\frac{\hat{p}_1}{m_1}\left(\rs{\hat{Q}_R}+\rs{\hat{Q}_L}\right)+\beta u_{4L}\la{4^L}\frac{\hat{P}}{m_P}\rs{\hat{Q}_R}\right)\nonumber\\
&=\delta^{(2)}\left((\alpha+\beta\gamma)u_{4L}\la{4^L}\frac{\hat{p}_1}{m_1}\rs{\hat{Q}_R}+\alpha u_{4L}\la{4^L}\frac{\hat{p}_1}{m_1}\rs{\hat{Q}_L}\right)\nonumber\\
\Rightarrow u_{4L}\la{4^L}\hat{p}_1\rs{\hat{Q}_R}&\sim\frac{-\alpha}{\alpha+\beta\gamma}u_{4L}\la{4^L}\hat{p}_1\rs{\hat{Q}_L}.
\end{align}
In the penultimate line, it has been used that 
\begin{align}
\la{q}\frac{\hat{P}}{m_P}=\gamma u_{4L}\la{4^L}\frac{\hat{p}_1}{m_1}+\lambda \la{q}\frac{\hat{p}_1}{m_1},
\end{align}
where the identity of the scalar $\lambda$ is unimportant and 
\begin{align}
\gamma=-\frac{\la{q}\hat{p}_1p_4\ra{q}}{m_Pm_1\da{q4^L}u_{4L}}.
\end{align}
Then 
\begin{align}
\delta^{(2)}\left(u_{4L}\la{4^L}\hat{p}_1\rs{\hat{Q}_R}\right)&=\delta^{(2)}\left(\frac{\alpha}{\beta\gamma}u_{4L}\la{4^L}\hat{p}_1\rs{\hat{Q}_R}+\frac{\alpha}{\beta\gamma}u_{4L}\la{4^L}\hat{p}_1\rs{\hat{Q}_R}\right)\nonumber\\
&=\delta^{(2)}\left(\left(\frac{\alpha}{\beta\gamma}\right)u_{4L}\la{4^L}\hat{p}_1\rs{Q}\right)
\end{align}
Thus $C^2=\left(\frac{\alpha}{\beta\gamma}\right)^2$ and the expression stated above may be obtained upon simplification through use of (\ref{M3PSK}), and the spin sums and the Weyl equations laid out in Appendix A of \cite{HKT:2018a}. 

The delta function $\delta^{(2)}\left(u_{4L}\la{4^L}\hat{p}_1\rs{Q}\right)$ may be amalgamated with the factor of \\ $\frac{1}{m_1^2\la{q}p_4\hat{p}_1\ra{q}}\delta^{(2)}\left(\la{q}\hat{p}_1\rs{Q}\right)$ derived above to give $\delta^{(4)}(Q)$.

Use of the alternative representation of the $3$-leg SUSY delta function in (\ref{3particle2}) may possibly be yield a simpler computation in this case once the Grassmann integral is performed (see \cite{Dennen:2009vk,Plefka:2014fta} for how this is done similarly in $6d$).

To derive (\ref{MagicPole}), using the spin sums and special massive kinematics (\ref{MassiveFactorize}) (and that $s_{\hat{1}\hat{2}}=s_{12}$ is unshifted),
\begin{align}
-u_{\hat{1}I}u_{\hat{1}L}s_{12}&=u_{\hat{1}I}u_{\hat{1}N}\epsilon_{ML}\left(\da{\hat{1}^N\hat{2}_J}\da{\hat{2}^J\hat{1}^M}+\ds{\hat{1}^N\hat{2}_J}\ds{\hat{2}^J\hat{1}^M}+\la{\hat{1}^N}\hat{p}_2\rs{\hat{1}^M}-\ls{\hat{1}^N}\hat{p}_2\ra{\hat{1}^M}\right)\nonumber\\
&=u_{\hat{1}I}u^{(L)}_{\hat{P}N}\epsilon_{ML}\left(\da{\hat{P}^N\hat{2}_J}\da{\hat{2}^J\hat{1}^M}+\ds{\hat{P}^N\hat{2}_J}\ds{\hat{2}^J\hat{1}^M}+\la{\hat{P}^N}\hat{p}_2\rs{\hat{1}^M}-\ls{\hat{P}^N}\hat{p}_2\ra{\hat{1}^M}\right)\nonumber\\
&=u_{\hat{1}I}u^{(L)}_{\hat{P}N}u^{(R)N}_{\hat{P}}u_{2J}\epsilon_{ML}\left(\ds{\hat{2}^J\hat{1}^M}-\da{\hat{2}^J\hat{1}^M}\right)\nonumber\\
&=u^{(L)}_{\hat{P}N}u^{(R)N}_{\hat{P}}\epsilon_{ML}u_{\hat{1}I}u^{(R)}_{\hat{P}J}\left(\ds{\hat{P}^J\hat{1}^M}-\da{\hat{P}^J\hat{1}^M}\right)\nonumber\\
&=u^{(L)}_{\hat{P}N}u^{(R)N}_{\hat{P}}u_{\hat{1}I}u^{(R)}_{\hat{P}J}u_{\hat{1}L}u^{(R)J}_{\hat{P}}=u_{\hat{1}I}u_{\hat{1}L}\left(u^{(L)}_{\hat{P}N}u^{(R)N}_{\hat{P}}\right)^2\nonumber\\
&\Rightarrow\left(u^{(L)}_{\hat{P}N}u^{(R)N}_{\hat{P}}\right)^2=-s_{12}.
\end{align}

\section{\texorpdfstring{$\mathcal{N}<4$}{N<4} SYM Superamplitudes from \texorpdfstring{$\mathcal{N}=4$}{N=4} SYM} \label{sec:projecting}

In this appendix we investigate how tree-level superamplitudes in Yang-Mills theories with less-than-maximal supersymmetry may be constructed from $\mathcal{N}=4$ SYM to determine the extent to which the valid BCFW shift may be exploited.

\subsection{Massless \texorpdfstring{$\mathcal{N}<4$}{N<4} SYM}

It was observed in \cite{Elvang:2011fx} that one may extract $\mathcal{N}=0,1,2$ submultiplets from the $\mathcal{N}=4$ massless vector multiplet via derivation or deletion of Grassmann variables. After defining extraction operators on states, one may then act with these operators on on-shell superamplitudes to find subamplitudes which describe the interactions of the submultiplets inside the $\mathcal{N}=4$ states. 

Subsequent to extraction, it is of interest to investigate whether the spectrum may be truncated in order to obtain superamplitudes of theories with fewer supersymmetries. In particular, we would like to know when it is possible for the states that have been removed from the external legs of the superamplitude by the extraction process to be omitted from the theory altogether while still retaining a meaningful superamplitude. Calling $S$ a set of extracted superfields closed under some $\mathcal{N}<4$ supersymmetries, we say that $S$ forms a `closed subsector' of the $\mathcal{N}=4$ tree-level theory if, for any tree-level subamplitude with external states only in $S$, it contains no contributions from off-shell states not in $S$. Then after extracting the subamplitudes of states in $S$, we may truncate the spectrum by ignoring the other states and we find the tree-level theory of states in $S$ enjoying some $\mathcal{N}<4$ supersymmetry. We denote the steps of extraction and truncation together as `projection' and say that this procedure projects from $\mathcal{N}=4$ SYM to the lower $\mathcal{N}$ theory.

The case discussed in \cite{Elvang:2011fx} is projection to pure (S)YM with $\mathcal{N}<4$. That is, one sets $S = \left\lbrace\mathcal{N}<4\,\text{vector multiplet}\right\rbrace$. The truncation is valid here because in any $\mathcal{N}$ (S)YM, all other states only couple to the vector multiplets in pairs, so lines of these particles in Feynman diagrams cannot be produced internally without closing in loops. So we may extract any tree-level amplitude in, for example, pure $\mathcal{N}=1$ SYM, $\mathcal{A}_n[G^\pm,G^\pm,\dots,G^\pm]_{\mathcal{N}=1}$ (where $G^\pm$ are massless $\mathcal{N}=1$ gluon superfields), from the tree-level amplitude $\mathcal{A}_n[G,G,\dots,G]_{\mathcal{N}=4}$ in a manner we will make precise momentarily.

The procedure for finding the extraction operators begins by first choosing which subset of supersymmetries our residual coherent states will be built out of. We then find the different ways one may take derivatives with respect to or delete the Grassmann variables corresponding to supersymmetries which will disappear. We end up manifestly with coherent states of the remaining supersymmetries. 

Consider first the extraction of $\mathcal{N}=2$ submultiplets which are coherent states of $Q_a$, $a = 1,2$ (thus reducing to a chiral superspace). For the massless multiplet, we may see on-shell the familiar statement that it consists of one $\mathcal{N}=2$ vector multiplet $G^\pm$ and one hypermultiplet $K$. With this choice of remaining supersymmetries, we may isolate these submultiplets from the massless $\mathcal{N}=4$ multiplet of (\ref{masslessmixed}) via
\begin{alignat}{2}\label{eqn:n=2chiral}
& G_{\mathcal{N}=2}^{+} =\half\frac{\partial}{\partial \eta^{\dagger m}}\frac{\partial}{\partial \eta^{\dagger}_m} G, \qquad &&G_{\mathcal{N}=2}^{-}=G\big|_{\eta^{\dagger}_m\rightarrow 0} \\
&K_{\mathcal{N}=2}^{+}= \frac{\partial}{\partial \eta^{\dagger}_3} G\big |_{\eta^{\dagger}_4\rightarrow 0}, \qquad &&\bar{K}_{\mathcal{N}=2}^{+}= \frac{\partial}{\partial \eta^{\dagger}_4} G \big|_{\eta^{\dagger}_3\rightarrow 0},
\end{alignat}
where the superscript refers to the helicity of the on-shell supermultiplet, and $K$ and $\bar{K}$ are the two $CP$-conjugate $\mathcal{N}=1$ chiral multiplets into which the $\mathcal{N}=2$ hypermultiplet may be decomposed. Here and throughout this appendix we use the $SU(2)\times SU(2)$ bronken $R$-symmetry notation for the superspace as in the first equation of (\ref{masslessmixed}).

If we were to instead extract the $\mathcal{N}=2$ submultiplets which are closed under $Q_1$ and $Q_3$, we would naturally end up in the non-chiral superspace for the massless multiplets
\begin{alignat}{2}\label{eqn:n=2nonchiral}
& G_{\mathcal{N}=2}^{+} = \frac{\partial}{\partial \eta^{\dagger}_4}  G \big |_{\eta^2\rightarrow 0}, \qquad &&G_{\mathcal{N}=2}^{-}=\frac{\partial}{\partial \eta^2}  G \big |_{\eta^{\dagger}_4\rightarrow 0} \\
&K_{\mathcal{N}=2} = G\big |_{\eta^2,\eta^{\dagger}_4\rightarrow 0}, \qquad &&\bar{K}_{\mathcal{N}=2} = \frac{\partial}{\partial \eta^{\dagger}_4} \frac{\partial}{\partial \eta^2} G.
\end{alignat}

We may also go further and extract the $\mathcal{N}=1$ submultiplets from $\mathcal{N}=4$, where we see the massless supermultiplet decompose into a vector multiplet (and $CP$ conjugate) $G^\pm$ and three chiral multiplets (and conjugate pairs) $\chi^{m}$ (for $m=2,3,4$) as
\begin{alignat}{2} \label{eqn:n=1massless}
&G^{+}_{\mathcal{N}=1} = \half \frac{\partial}{\partial \eta^{\dagger m}}\frac{\partial}{\partial \eta^{\dagger}_m} G\big|_{\eta^{2}\rightarrow 0}, \qquad 
&&G^{-}_{\mathcal{N}=1} = \frac{\partial}{\partial \eta^{2}} G\big|_{\eta^{\dagger}_m\rightarrow 0} \\
&\chi^{2+}_{\mathcal{N}=1} = \half \frac{\partial}{\partial \eta^{\dagger m}}\frac{\partial}{\partial \eta^{\dagger}_m}  \frac{\partial}{\partial \eta^{2}} G, \qquad 
&&\chi^{2-}_{\mathcal{N}=1} = G\big|_{\eta^{\dagger}_3,\eta^{\dagger}_4,\eta^2\rightarrow 0}, \\
&\chi^{m+}_{\mathcal{N}=1} =\frac{\partial}{\partial \eta^{\dagger}_m} G\big|_{\eta^{\dagger m},\eta^{2}\rightarrow 0}, \qquad 
&&\chi^{m-}_{\mathcal{N}=1} = \frac{\partial}{\partial \eta^{\dagger m}}  \frac{\partial}{\partial \eta^{2}} G\big|_{\eta^{\dagger}_m\rightarrow 0} 
\end{alignat}
where $m$ indexes which $R$-index we took a derivative with respect to, which is merely a more compact notation than we used for the massless hypermultiplet above.

Since the $\mathcal{N}=2$ vector multiplet forms a closed subsector in pure SYM on its own, it can be split up into $\mathcal{N}=1$ submultiplets $G^\pm$, $\chi^{4\pm}$. As discussed in \cite{Dixon:2010ik}, one may find amplitudes for fundamental quarks in ($\mathcal{N}=0$) QCD from color-ordered amplitudes of adjoint gluinos merely by using different color factors when summing over color-orderings. Likewise, one may study $\mathcal{N}=1$ SQCD with one flavor of massless fundamental quark chiral superfield at tree-level using this construction.

We can then proceed even further and go to $\mathcal{N}=0$ non-supersymmetric Yang-Mills by simply considering each component field separately. As above, one may find closed subsectors from $\mathcal{N}=1,2$ supersymmetry which include, in addition to the gluons, massless fermions (from $\mathcal{N}=1$) or both massless fermions and scalars (from $\mathcal{N}=2$). Special combinations of amplitudes in the projected $\mathcal{N}=2$ SYM theory were used by \cite{Dixon:2010ik} in order to compute tree QCD amplitudes with multiple quark flavours while avoiding internal off-shell interactions with their scalar partners.

\subsection{Massive \texorpdfstring{$\mathcal{N}<4$}{N<4} SYM}

The next question is whether any of this structure survives on the $\mathcal{N}=4$ Coulomb branch now that we have another type of multiplet. We will not, in fact, find closed subsectors which include massive states, for the reason that any massive submultiplet couples at tree-level to all of the massless submultiplets, which will be apparent after we give the extraction operators for massive states. However, we will be able to find effectively closed subsectors by restricting our attention to certain subsets of (super)amplitudes, in which only states in that subsector appear internally. This will allow us to deduce some interesting features of various $\mathcal{N}<4$ theories at tree-level.

If we extract $\mathcal{N}=2$ coherent states of the $Q_1$ and $Q_2$ supersymmetries, then for the massive multiplet we are left with the long vector multiplet of $\mathcal{N}=2$. This has exactly the same field content as the short $\mathcal{N}=4$ multiplet with which we've been working. It's clear from the form of the central charge that restricting our attention to the supercharges which anticommute leaves a massive multiplet without a central charge. We can then extract tree-level superamplitudes for $\mathcal{N}=2$ by simply extracting the $\mathcal{N}=2$ massless submultiplets in the chiral superspace as in (\ref{eqn:n=2chiral}). One finds nonzero tree-level three-leg amplitudes of the massive $\mathcal{N}=2$ vector (denoted as $\Omega$ here) with both the massless vector and the massless hypermultiplet, for example, through
\begin{equation}
\begin{split}
\mathcal{A}_{\mathcal{N}=2}[\Omega,G^{+},\overline{\Omega}]&=\half \frac{\partial}{\partial \eta^{\dagger m}_2}\frac{\partial}{\partial \eta^{\dagger}_{2m}} \mathcal{A}[\mathcal{W},G,\overline{\mathcal{W}}] \\
\mathcal{A}_{\mathcal{N}=2}[\Omega,K^+,\overline{\Omega}]&=\frac{\partial}{\partial \eta^{\dagger 3}_2}\mathcal{A}[\mathcal{W},G,\overline{\mathcal{W}}]\big|_{\eta_{2}^{\dagger 4}\rightarrow 0}.
\end{split}
\end{equation}
We thus cannot truncate the spectrum by deleting the hypermultiplets, as these appear in factorization channels of higher-leg subamplitudes containing only external massless and massive vectors.

Our other option to obtain $\mathcal{N}=2$ submultiplets is to extract coherent states of a pair of supersymmetries whose supercharges have nonzero anticommutator, for example $Q_1, Q_3$. For the massless multiplets, this puts us in the non-chiral superspace representation of (\ref{eqn:n=2nonchiral}). For the massive multiplets, this extracts the BPS multiplets of $\mathcal{N}=2$, which are simply the massive $\mathcal{N}=1$ supermultiplets (the extraction of which will be demonstrated next). The massive states may be described solely as coherent states of $Q_1$ in both $\mathcal{N}=1$ and short $\mathcal{N}=2$ cases, so the differences between subamplitudes with either BPS $\mathcal{N}=2$ submultiplets or massive $\mathcal{N}=1$ submultiplets are attributable to the massless states present (\ref{eqn:n=1massless}).

The massive multiplets decompose into one $\mathcal{N} = 1$ vector multiplet and two $\mathcal{N} = 1$ chiral multiplets as
\begin{alignat}{3}
\mathcal{Q}_{\mathcal{N}=1} &= \half \frac{\partial}{\partial \eta^2_{I}}\frac{\partial}{\partial \eta^{2I}} \mathcal{W} \qquad \mathcal{W}^I_{\mathcal{N}=1} &= \frac{\partial}{\partial \eta^2_{I}} \mathcal{W}\big|_{\eta^{2}_J\rightarrow 0} \qquad 
\mathcal{Q}'_{\mathcal{N}=1} &= \mathcal{W}\big|_{\eta^{2}_I\rightarrow 0} \\
\overline{\mathcal{Q}}_{\mathcal{N}=1} &= \overline{\mathcal{W}}\big|_{\eta^2_{I}\rightarrow 0} \qquad\qquad \overline{\mathcal{W}}^I_{\mathcal{N}=1} &= \frac{\partial}{\partial \eta^2_{I}} \overline{\mathcal{W}}\big|_{\eta^2_J\rightarrow 0} \qquad 
\overline{\mathcal{Q}'}_{\mathcal{N}=1} &= \half \frac{\partial}{\partial \eta^2_{I}}\frac{\partial}{\partial \eta^{2I}} \overline{\mathcal{W}}.
\end{alignat}
The massive matter states may be alternatively grouped into massive $\mathcal{N}=2$ hypermultiplets, just as for the massless case above. To reiterate, we interpret these either as $\mathcal{N}=1$ or $\mathcal{N}=2$ submultiplets depending upon which massless states are in the amplitude, which, at this point in the discussion, is simply a collection of components of a $\mathcal{N}=4$ superamplitude. We could of course go further and extract the $\mathcal{N}=0$ components easily.

Now that we have all of the extraction operators, we may ask which tree-level amplitudes may be obtained by truncating the spectrum. While we cannot project from the Coulomb branch to an entire theory of massive $\mathcal{N}<4$ SYM, we may still be able to project onto particular amplitudes in $\mathcal{N}<4$ SYM theories. The simplest examples are the three-leg amplitudes of any minimally-coupled matter with Yang-Mills theory. The extraction of the $\mathcal{N}=1$ three-leg amplitude for two equal mass vector superfields and a positive-helicity massless vector gives
\begin{align}
\mathcal{A}[\mathcal{W}^I,\overline{\mathcal{W}}^J,G^+] &= \frac{\partial}{\partial\eta^2_{1I}} \frac{\partial}{\partial\eta^2_{2J}} \half \frac{\partial}{\partial \eta_3^{\dagger m}}\frac{\partial}{\partial \eta^{\dagger}_{3m}} \mathcal{A}[\mathcal{W},\overline{\mathcal{W}},G]\big|_{\eta^2_{1K},\eta^2_{2K},\eta^2_{3}\rightarrow 0} \nonumber\\
&= \delta^{(2)}(Q^\dagger) g \ls{1^I}^\alpha \ls{2^J}^\beta \left( \frac{1}{m x} \epsilon_{\alpha\beta} - \frac{1}{m^2} \rs{3}_\alpha \rs{3}_\beta \right).
\end{align}
By comparison with the discussion in \cite{HKT:2018a}, we see that at tree-level the anomalous magnetic dipole moment of the massive vector superfield has been set to zero.

We may next look for higher leg tree-level amplitudes that are not affected by the absence of truncated particles. The key is that the massive states couple in pairs to the massless states in $\mathcal{N}=4$, so this property is inherited in each projected theory and, as above, the other massless multiplets also couple to the reduced supersymmetry massless vector multiplet in pairs. This allows us to argue, for example, that the 2 massive leg, $n-2$ massless vector superamplitudes $\mathcal{A}_n[M,\overline{M},G^\pm,G^\pm,\dots,G^\pm]$ (gluon helicities arbitrary) may be found via an appropriate projection, where $M$ may here be any of the massive multiplets of $\mathcal{N}<4$. From the above, no other states may appear internally. Of course this can also be taken one step further down to $\mathcal{N}=0$, which allows us to find the tree-level amplitudes for any number of gluons and two massive particles of any spin $\leq 1$. 

Furthermore, this projection allows us to see an interesting feature for the $\mathcal{N}=1$ all-plus-helicity amplitudes. The $\mathcal{N}=4$ Coulomb branch amplitudes have Grassmann degree $2n$, while the extraction operators for such an $\mathcal{N}=1$ amplitude involve $2(n-2)+2$ derivatives, so that these subamplitudes will have Grassmann degree $2$. This means that the Grassmann delta function saturates the Grassmann dependence, so these tree-level superamplitudes may be entirely characterized once one component amplitude is known, for example
\begin{equation}
\mathcal{A}_n[\mathcal{Q},\overline{\mathcal{Q}},G^+,G^+,\dots,G^+] = \frac{-1}{m} \delta^{(2)}(Q^\dagger) A_n[\widetilde{Q}_R,\overline{\widetilde{Q}}_R,g^+,g^+,\dots,g^+],
\end{equation}
in the notation of Section 5.2 of \cite{HKT:2018a}. This means that we may perform massless $\mathcal{N}=0$ BCFW recursion to find a single component amplitude and get the rest for free, rather than needing to perform the recursion in $\mathcal{N}=4$ and then project down. In particular, we may upgrade already-known results for all-$n$ amplitudes in QCD \cite{Badger:2005zh,Schwinn:2006ca, Schwinn:2007ee,Ochirov:2018uyq} to full $\mathcal{N}=1$ SQCD superamplitudes.

Some simple examples of tree-level amplitudes that may be obtained by projection to $\mathcal{N}=1$ SYM with massive vectors are 
\begin{align}
\mathcal{A}[\mathcal{W}^I,\overline{\mathcal{W}}^J,G^+,G^+] &= \frac{\delta^{(2)}(Q^\dagger) \da{1^I 2^J} \ds{34}^2}{(p_1+p_2)^2 \left( (p_2 + p_3)^2 + m^2\right)},\\
\mathcal{A}[\mathcal{W}^I,\overline{\mathcal{W}}^J,G^+,G^-] &= \frac{\delta^{(2)}(Q^\dagger) \left( \da{1^I 4}\ds{2^J 3} + \da{2^J 4}\ds{1^I 3}\right)\left( \ds{1^K 3}\eta_{1K} - \ds{2^L 3} \eta_{2L}\right)}{(p_1+p_2)^2 \left( (p_2 + p_3)^2 + m^2\right)}.
\end{align}

The above is merely an initial exploration into what the Coulomb branch can tell us about massive amplitudes in Yang-Mills theories with fewer supersymmetries.

\bibliography{superspace}{}
\bibliographystyle{JHEP}

\end{document}